\documentclass[usenames, dvipsnames]{article} 
\usepackage[nonatbib, final]{neurips_2022}

\usepackage{amsmath,amsfonts,bm}

\def\eqref#1{equation~\ref{#1}}

\def\1{\bm{1}}

\DeclareMathAlphabet{\mathsfit}{\encodingdefault}{\sfdefault}{m}{sl}
\SetMathAlphabet{\mathsfit}{bold}{\encodingdefault}{\sfdefault}{bx}{n}

\usepackage{hyperref}
\usepackage{url}
\hypersetup{
  colorlinks,
  citecolor=RoyalPurple,
  linkcolor=Magenta,
  urlcolor=Blue}
\usepackage[utf8]{inputenc} 
\usepackage{url}            
\usepackage{booktabs}       
\usepackage{amsfonts}       
\usepackage[scaled]{beramono} 
\usepackage[T1]{fontenc}    
\usepackage{nicefrac}       
\usepackage{microtype}      
\usepackage{xcolor}         
\usepackage{bm}             
\usepackage{amsmath}        
\usepackage{mathtools}      
\usepackage[linesnumbered,ruled]{algorithm2e}

\SetCommentSty{algocommfont}

\SetKwSty{algokwfont}

\SetFuncSty{algofnfont}
\usepackage{multicol}       
\usepackage{graphicx}       
\usepackage{diagbox}        
\usepackage{multirow}       
\usepackage{wrapfig}        
\usepackage{upgreek}        
\usepackage[compact]{titlesec} 

\title{FourierNets enable the design of highly non-local optical encoders for computational imaging}

\author{
  Diptodip Deb$^1$
  \qquad
  Zhenfei Jiao$^{1,2}$
  \qquad
  Ruth Sims$^{1,3}$
  \qquad
  Alex B.~Chen$^{1,4}$
  \And
  Michael Broxton$^{5,\dagger}$
  \qquad
  Misha B.~Ahrens$^{1,\dagger}$
  \qquad
  Kaspar Podgorski$^{1,\dagger}$
  \qquad
  Srinivas C.~Turaga$^{1,\dagger}$\\\\
  $^1$ HHMI Janelia Research Campus, $^2$ Wuhan National Laboratory for Optoelectronics, HUST\\
  $^3$ Institut de la Vision, Sorbonne Université, CNRS, $^4$ Harvard University, $^5$ Stanford University,\\
  \texttt{\{debd,simsr,ahrensm,podgorskik,turagas\}@janelia.hhmi.org} \\
  \texttt{zfjiao@hust.edu.cn}\\
  \texttt{abchen@g.harvard.edu}\\
  \texttt{broxton@alumni.stanford.edu}\\
  {\small $^\dagger$ equal contribution}
}

\begin{document}

\maketitle

\begin{abstract}
Differentiable simulations of optical systems can be combined with deep learning-based reconstruction networks to enable high performance computational imaging via end-to-end (E2E) optimization of both the optical encoder and the deep decoder. This has enabled imaging applications such as 3D localization microscopy, depth estimation, and lensless photography via the optimization of local optical encoders. More challenging computational imaging applications, such as 3D snapshot microscopy which compresses 3D volumes into single 2D images, require a highly non-local optical encoder. We show that existing deep network decoders have a locality bias which prevents the optimization of such highly non-local optical encoders. We address this with a decoder based on a shallow neural network architecture using global kernel Fourier convolutional neural networks (FourierNets). We show that FourierNets surpass existing deep network based decoders at reconstructing photographs captured by the highly non-local DiffuserCam optical encoder. Further, we show that FourierNets enable E2E optimization of highly non-local optical encoders for 3D snapshot microscopy. By combining FourierNets with a large-scale multi-GPU differentiable optical simulation, we are able to optimize non-local optical encoders 170$\times$ to 7372$\times$ larger than prior state of the art, and demonstrate the potential for ROI-type specific optical encoding with a programmable microscope.
\end{abstract}

\section{Introduction}
\label{introduction}
Modern computational optics relies on the end-to-end (E2E) optimization of the optical system (optical encoder) and the computational image reconstruction algorithm (computational decoder). This has been enabled by the development of differentiable physics-based simulations of optical encoders, paired with deep network based decoders. This E2E approach has been successfully applied to the estimation of depth from defocus \cite{Chang_Wetzstein_2019, Dun_Dun_Ikoma_Wetzstein_Wang_Wang_Cheng_Cheng_Cheng_Peng_et_al_2020, Ikoma_Nguyen_Metzler_Peng_Wetzstein_2021, Wu_Boominathan_Chen_Sankaranarayanan_Veeraraghavan_2019}, lensless photography \cite{Hua_Nakamura_Asif_Sankaranarayanan_2020}, and particle localization microscopy \cite{Nehme_Freedman_Gordon_Ferdman_Weiss_Alalouf_Naor_Orange_Michaeli_Shechtman_2020, Ikoma_Kudo_Peng_Broxton_Wetzstein_2021}. Despite this wide success, only the space of local optical encoders has been explored in this fashion due to the computational expense of physics-based simulation of highly non-local optical encoders \cite{Ikoma_Nguyen_Metzler_Peng_Wetzstein_2021}. This is particularly the case for 3D snapshot microscopy which is well-known to require highly non-local optical encoders, and for which no E2E solutions have yet been demonstrated.

We show that there are actually two issues blocking the E2E optimization of highly non-local optical encoders. First, the computational expense of physics-based simulation of non-local optical encoders is significantly greater, since they are orders of magnitude larger than local encoders. And second, the locality bias inherent to deep network architectures currently used for computational imaging prevents decoding from non-local solutions. In this paper, we primarily focus on developing a computational framework for engineering highly non-local optical encoders by solving both issues. We solve the first issue by developing a large-scale multi-GPU differentiable optical simulation, and the second by developing a new FourierNet architecture based deep network decoder.

In 3D snapshot microscopy, all existing methods \cite{Levoy_Ng_Adams_Footer_Horowitz_2006, Cong_Wang_Chai_Hang_Shang_Yang_Bai_Du_Wang_Wen_2017, Linda_Liu_Kuo_Antipa_Yanny_Waller_2020, Pegard_Liu_Antipa_Gerlock_Adesnik_Waller_2016, Prevedel_Yoon_Hoffmann_Pak_Wetzstein_Kato_Schrodel_Raskar_Zimmer_Boyden_et_al_2014, Skocek_Nobauer_Weilguny_Martinez_Traub_Xia_Molodtsov_Grama_Yamagata_Aharoni_Cox_et_al_2018, Yanny_Antipa_Ng_Waller_Waller_2019, Wagner_Beuttenmueller_Norlin_Gierten_Boffi_Wittbrodt_Weigert_Hufnagel_Prevedel_Kreshuk_2021, Antipa_Kuo_Heckel_Mildenhall_Bostan_Ng_Waller_2018, Liu_Madhavan_Antipa_Kuo_Kato_Waller_2019, Asif_Ayremlou_Veeraraghavan_Baraniuk_Sankaranarayanan_2015, Hua_Nakamura_Asif_Sankaranarayanan_2020, Yanny_Antipa_Liberti_Dehaeck_Monakhova_Liu_Shen_Ng_Waller_2020} employ a highly non-local optical encoding to optically transform a 3D volume into a single 2D camera image, which is computationally decoded to reconstruct the 3D volume. Fast volumetric imaging is invaluable across biology, including for imaging neural activity across the whole brain of an animal \cite{Ahrens_Orger_Robson_Li_Keller_2013, Mu_Bennett_Rubinov_Narayan_Yang_Tanimoto_Mensh_Looger_Ahrens_2019, Yang_Yuste_2017}. Here, we focus on developing a 3D snapshot microscope for imaging neuronal activity across the whole brain of the larval zebrafish (\textit{Danio rerio}) at camera rates exceeding 100Hz. This is two orders of magnitude faster than the fastest conventional microscopes --- light sheet imaging of whole brain neural activity can only achieve 0.5Hz - 2Hz volume rates \cite{Ahrens_Orger_Robson_Li_Keller_2013}. This improvement of temporal resolution is essential for imaging with fast calcium indicators \cite{Zhang_Rozsa_Liang_Bushey_Wei_Zheng_Reep_Broussard_Tsang_Tsegaye_et_al_2021} and voltage indicators \cite{Abdelfattah_Kawashima_Singh_Novak_Liu_Shuai_Huang_Campagnola_Seeman_Yu_et_al_2019}.

Newly developed programmable microscopes with up to $10^6$ free parameters, e.g. pixels on a spatial light modulator (SLM), enable the implementation of a rich space of optical encodings, and present the possibility of direct optimization of snapshot microscope parameters specifically for particular ROI types and imaging tasks. While our paper primarily focuses on addressing the computational challenges of E2E optimization of highly non-local optical encoders, we also demonstrate proof-of-concept that an SLM-based programmable microscope can implement such an engineered non-local optical encoder, paving the way to implementing ROI-type and task specific optical encoders in a single physical microscope. In contrast, 3D snapshot imaging has classically been performed using fixed optical encoders based on microlens arrays \cite{Levoy_Ng_Adams_Footer_Horowitz_2006, Cong_Wang_Chai_Hang_Shang_Yang_Bai_Du_Wang_Wen_2017, Linda_Liu_Kuo_Antipa_Yanny_Waller_2020, Pegard_Liu_Antipa_Gerlock_Adesnik_Waller_2016, Prevedel_Yoon_Hoffmann_Pak_Wetzstein_Kato_Schrodel_Raskar_Zimmer_Boyden_et_al_2014, Skocek_Nobauer_Weilguny_Martinez_Traub_Xia_Molodtsov_Grama_Yamagata_Aharoni_Cox_et_al_2018, Yanny_Antipa_Ng_Waller_Waller_2019, Wagner_Beuttenmueller_Norlin_Gierten_Boffi_Wittbrodt_Weigert_Hufnagel_Prevedel_Kreshuk_2021}, pseudorandom diffusers/masks \cite{Antipa_Kuo_Heckel_Mildenhall_Bostan_Ng_Waller_2018, Liu_Madhavan_Antipa_Kuo_Kato_Waller_2019}, and designed or optimized diffusers/masks \cite{Asif_Ayremlou_Veeraraghavan_Baraniuk_Sankaranarayanan_2015, Hua_Nakamura_Asif_Sankaranarayanan_2020, Yanny_Antipa_Liberti_Dehaeck_Monakhova_Liu_Shen_Ng_Waller_2020}, with some of these methods using deep learning for reconstruction \cite{Pegard_Liu_Antipa_Gerlock_Adesnik_Waller_2016, Wagner_Beuttenmueller_Norlin_Gierten_Boffi_Wittbrodt_Weigert_Hufnagel_Prevedel_Kreshuk_2021}.

\textbf{Problem statement}
We define an optical encoder as $\mathbf{M}_{\bm{\upphi}}$ parameterized by $\bm{\upphi}$ and a computational decoder as $\mathbf{R}_{\bm{\uptheta}}$ parameterized by $\bm{\uptheta}$. Optical encoders can usually be modeled as linear transfer functions implemented by the optics. In special cases, such as for the optical systems explored in this paper, they can be represented as a linear convolution filter called a ``point spread function'' which has its coefficients computed by a wave optics simulation, dependent on the physical parameters $\bm{\upphi}$ of the optical system (Appendix \ref{appendixforwardsim}). We wish to develop a reconstruction network architecture for decoding non-local encodings in 2D images $\mathbf{c}$ produced by $\mathbf{M}_{\bm{\upphi}}$ and also to enable the end-to-end optimization of $\bm{\upphi}$ to produce non-local encodings. We'll consider two applications to investigate these optimization problems: (1) lensless photography (Figure \ref{figdlmdcomparison}), where we optimize only the reconstruction network $\mathbf{R}_{\bm{\uptheta}}$ to reconstruct images of natural scenes from camera images captured by the DiffuserCam \cite{Monakhova_Yurtsever_Kuo_Antipa_Yanny_Waller_2019}, and (2) 3D snapshot microscopy using an SLM-based programmable microscope (Figure \ref{figoverview}A), where we perform E2E optimization of both the reconstruction network $\mathbf{R}_{\bm{\uptheta}}$ and the optical encoding $\mathbf{M}_{\bm{\upphi}}$ based on the SLM parameters in order to image 3D volumes $\mathbf{v}$ and reconstruct 3D volumes $\mathbf{\hat{v}}$. While we focus on solving the computational challenges involved in this E2E optimization of a large highly non-local optical encoder, we also demonstrate a proof-of-concept implementation of our optimized optical encoder using an SLM-based programmable microscope.

\subsection{Our contributions}
\begin{enumerate}
    \item We developed a large-scale, parallel, multi-GPU differentiable wave optics simulation of a programmable microscope, based on a $4f$ optical model with a phase mask ($\bm{\upphi}$) implemented using a spatial light modulator (SLM), described further in Appendix \ref{appendixforwardsim}. SLMs ($\bm{\upphi}$) can have over $10^6$ optimizable parameters, which we can feasibly simulate and optimize to produce PSFs with 170$\times$ to 7372$\times$ more unique voxels than previous attempts at deep learning PSF optimization using single GPUs \cite{Nehme_Freedman_Gordon_Ferdman_Weiss_Alalouf_Naor_Orange_Michaeli_Shechtman_2020, Ikoma_Nguyen_Metzler_Peng_Wetzstein_2021} (Appendix \ref{appendixvoxelcomparison}).
    \item We collected a large dataset of high resolution 3D confocal volumes of zebrafish larvae for the purpose of ROI-type specific end-to-end optimization of optical encoders.
    \item We introduce an efficient FourierNet reconstruction network architecture for decoding from non-local optical encoders using very large global convolutions implemented via Fourier convolutions.
    \item We show that our networks outperform the state-of-the-art deep decoders for DiffuserCam based lensless photography \cite{Monakhova_Yurtsever_Kuo_Antipa_Yanny_Waller_2019} and for 3D snapshot microscopy.
    \item Our method enables, for the first time, direct end-to-end optimization of highly non-local optical encoders in the space of spatial light modulator (SLM) pixels with over $10^6$ parameters. In simulation, we demonstrate the potential for significant improvements in imaging resulting from ROI-type specific optimization of optical encoders.
\end{enumerate}

\subsection{Prior work}
Neural network architectures for computational imaging have all used convolution layers with small filters. End-to-end optimization of optical encoders have largely been performed with UNet-based architectures \cite{Chang_Wetzstein_2019, Dun_Dun_Ikoma_Wetzstein_Wang_Wang_Cheng_Cheng_Cheng_Peng_et_al_2020, Ikoma_Nguyen_Metzler_Peng_Wetzstein_2021, Wu_Boominathan_Chen_Sankaranarayanan_Veeraraghavan_2019, Hua_Nakamura_Asif_Sankaranarayanan_2020, Ikoma_Kudo_Peng_Broxton_Wetzstein_2021}, with one method using a ResNet-based architecture \cite{Nehme_Freedman_Gordon_Ferdman_Weiss_Alalouf_Naor_Orange_Michaeli_Shechtman_2020}. Such optimization has always led to local optical encodings.
End-to-end optimization has never been attempted for large-field of view 3D snapshot microscopy due to the difficulty of simulating and reconstructing from non-local encoders.
However, small filter convolutional deep networks have been used in a non-end-to-end manner to reconstruct volumes from 3D snapshot microscopes designed using microlens arrays \cite{Wagner_Beuttenmueller_Norlin_Gierten_Boffi_Wittbrodt_Weigert_Hufnagel_Prevedel_Kreshuk_2021, Wang_Zhu_Zhang_Li_Yi_Li_Yang_Ding_Zhen_Gao_et_al_2021, Yanny_Yanny_Monakhova_Monakhova_Shuai_Waller_2022}. One recent hybrid approach combines a UNet with a differentiable approximate inverse method (Wiener filter) to handle non-local spatially varying (non-convolutional) optical encoders \cite{Yanny_Yanny_Monakhova_Monakhova_Shuai_Waller_2022}.
For photography and MRI, another hybrid approach of deep learning combined with unrolling iterations of traditional deconvolution algorithms provides the benefits of fast amortized optimization and higher quality reconstructions due to learning of structural priors \cite{Diamond_Sitzmann_Heide_Wetzstein_2017, Dong_Roth_Schiele_2020, Monakhova_Yurtsever_Kuo_Antipa_Yanny_Waller_2019, Wang_Zhu_Zhang_Li_Yi_Li_Yang_Ding_Zhen_Gao_et_al_2021}. We note that these hybrid approaches typically require measurement of the optical encoder of the system, whereas our method does not and can learn to produce high quality reconstructions using only pairs of ground truth and system images.

In our work, we demonstrate that convolution layers with large filters implemented efficiently in the Fourier domain enable the end-to-end learning of highly non-local optical encoders for 3D snapshot microscopy. Large convolution filters have been shown to be helpful for other computer vision applications such as semantic segmentation and salient object detection  \cite{Peng_Zhang_Yu_Luo_Sun_2017, Mu_Xu_Zhang_2019}, and the Fourier domain parameterization of small filters has been described previously \cite{Rippel_Snoek_Adams_2015}.

A pioneering strategy in 3D snapshot microscopy has been light field microscopy \cite{Levoy_Ng_Adams_Footer_Horowitz_2006}, which employs a microlens array at the microscope's image plane to create subimages encoding both amplitude and phase of light \cite{Levoy_Ng_Adams_Footer_Horowitz_2006, Adelson_Wang_1992}. A variety of microlens-array-based light field microscopes have been used to perform whole-brain imaging \cite{Levoy_Ng_Adams_Footer_Horowitz_2006, Pegard_Liu_Antipa_Gerlock_Adesnik_Waller_2016, Yanny_Antipa_Ng_Waller_Waller_2019, Yang_Yuste_2017, Skocek_Nobauer_Weilguny_Martinez_Traub_Xia_Molodtsov_Grama_Yamagata_Aharoni_Cox_et_al_2018, Prevedel_Yoon_Hoffmann_Pak_Wetzstein_Kato_Schrodel_Raskar_Zimmer_Boyden_et_al_2014, Cong_Wang_Chai_Hang_Shang_Yang_Bai_Du_Wang_Wen_2017, Grosenick_Broxton_Kim_Liston_Poole_Yang_Andalman_Scharff_Cohen_Yizhar_et_al_2017, Yanny_Antipa_Liberti_Dehaeck_Monakhova_Liu_Shen_Ng_Waller_2020}. \cite{Yanny_Antipa_Liberti_Dehaeck_Monakhova_Liu_Shen_Ng_Waller_2020} optimizes the placement of microlenses, but not in an end-to-end manner. Despite variation in design, microlens-based microscopes have, to various degrees, four main limitations that can be improved: 1) blocking or scattering of light between microlenses, causing light inefficiency, 2) not making use of all pixels on the camera to encode a 3D volume, leading to inefficient compression and suboptimal reconstructions $\mathbf{\hat{v}}$, 3) aliasing at some planes with generally nonuniform axial resolution, and 4) a fixed optical encoding scheme.

An alternative to using microlenses is to implement a coded detection strategy using a phase mask or diffuser to spread light broadly across the camera sensor \cite{Antipa_Kuo_Heckel_Mildenhall_Bostan_Ng_Waller_2018, Asif_Ayremlou_Veeraraghavan_Baraniuk_Sankaranarayanan_2015, Broxton_2017, Hua_Nakamura_Asif_Sankaranarayanan_2020, Linda_Liu_Kuo_Antipa_Yanny_Waller_2020, Liu_Madhavan_Antipa_Kuo_Kato_Waller_2019, Yanny_Antipa_Liberti_Dehaeck_Monakhova_Liu_Shen_Ng_Waller_2020}. The designed phase masks can be implemented either by manufacturing a custom optical element or using a programmable SLM \cite{Dun_Dun_Ikoma_Wetzstein_Wang_Wang_Cheng_Cheng_Cheng_Peng_et_al_2020, Ikoma_Nguyen_Metzler_Peng_Wetzstein_2021, Wu_Boominathan_Chen_Sankaranarayanan_Veeraraghavan_2019, Hua_Nakamura_Asif_Sankaranarayanan_2020, Yanny_Antipa_Liberti_Dehaeck_Monakhova_Liu_Shen_Ng_Waller_2020, Nehme_Freedman_Gordon_Ferdman_Weiss_Alalouf_Naor_Orange_Michaeli_Shechtman_2020}. Using a programmable element allows different microscope parameters to be used for different sample and ROI types.


\titlespacing{\section}{0pt}{0ex}{0ex}
\section{Methods}
\label{methods}
\begin{figure}[h]
  \includegraphics[width=\linewidth]{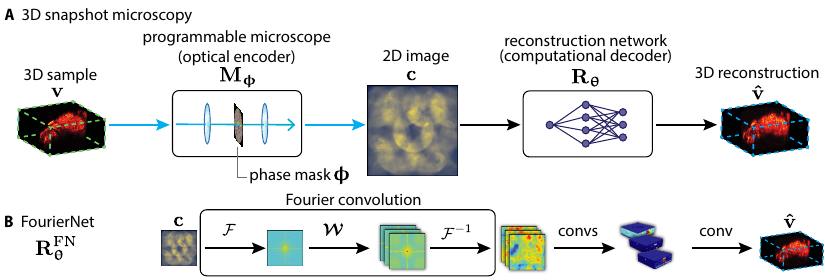}
  \caption{Overview of our problem setup and our proposed network architectures. Top row (\textbf{A}) shows the problem of 3D snapshot microscopy, where we computationally reconstruct a 3D volume from a 2D image. Bottom row (\textbf{B}) shows our proposed FourierNet architecture, which includes a \textbf{Fourier convolution} layer that enables efficient computation of global features.}
  \label{figoverview}
  \vspace{-0.8cm}
\end{figure}

We show our network architecture and an overview of autoencoder training both the microscope parameters $\bm{\upphi}$ and reconstruction network parameters $\bm{\uptheta}$ in Figure \ref{figoverview}A. The programmable microscope is simulated by a differentiable implementation of a wave-optics model of light propagation. We have selected a programmable microscope design based on pupil-plane phase modulation with a programmable spatial light modulator, for which imaging is well-approximated by a computationally-efficient convolution \cite{Goodman_2017}. A detailed description of our simulation is provided in Appendix \ref{appendixforwardsim}. For lensless photography, there is no optical simulation because the images have been collected on a real camera.
\titlespacing{\subsection}{0pt}{1ex}{0ex}

\subsection{FourierNet for decoding non-local optical encoders}
\label{fourierconv}
Because images created by optical encoders can potentially encode signals from the incoming light field to any location in the camera image in a non-local manner, it is essential that a reconstruction network have global context. Existing multi-scale architectures such as the UNet \cite{Ronneberger_Fischer_Brox_2015} can achieve global context, but at the expense of many computation layers with small convolution kernels which we speculate to have a local information bias which is inappropriate for computational optics. In this paper, we introduce relatively shallow architectures for computational imaging, which rely on convolution layers with very large filters, implemented efficiently in the Fourier domain \cite{Rippel_Snoek_Adams_2015}.  

\textbf{FourierNet} We propose a simple three layer convolutional network architecture with very large global convolutions at the very first layer, followed by two standard local convolutional layers (Figure \ref{figoverview}B). We define a global convolution as a convolution with kernel size equal to the input image. Such a convolution achieves global context in a single step but is computationally expensive. Global convolutions are implemented more efficiently in the Fourier domain, yielding a speed up of two orders of magnitude. Due to the use of Fourier convolutions to enable global context, we call our architecture FourierNet. In contrast to a typical UNet which can contain many tens of convolution layers, the FourierNet is only three layers deep, which requires backpropagation through fewer layers compared to a typical UNet with the same receptive field.

\textbf{Fourier domain convolutions}
It is well-known that large kernel convolutions can be implemented more efficiently in the Fourier domain \cite{Oppenheim_Willsky_2013, Mathieu_Henaff_LeCun_2014, Vasilache_Johnson_Mathieu_Chintala_Piantino_LeCun_2015}. A naive implementation of global convolution requires $\mathcal{O}(N^2)$ operations, where $N$ is number of pixels in both the image and the kernel. An alternative global convolution implementation is to Fourier transform $\mathcal{F}$ the input $\mathbf{x}$ and convolution kernel $\mathbf{w}$, perform element-wise multiplication in Fourier space, and finally inverse Fourier transform $\mathcal{F}^{-1}$, requiring only $\mathcal{O}(N\log{N})$ operations \cite{Mathieu_Henaff_LeCun_2014, Vasilache_Johnson_Mathieu_Chintala_Piantino_LeCun_2015}. Following \cite{Rippel_Snoek_Adams_2015}, we store and optimize the weights in Fourier space $\bm{\mathcal{W}}$. This over-parameterization costs $8\times$ the memory of an equivalent real valued large filter but saves the computational cost of Fourier transforming the real-valued weights (Appendix \ref{appendiximplementationdetails}). Thus a Fourier convolution is defined:
\setlength{\abovedisplayskip}{3.5pt}
\setlength{\belowdisplayskip}{\abovedisplayskip}
\begin{equation}
  \mathbf{Re}\{\mathcal{F}^{-1}\left\{ \bm{\mathcal{W}} \odot \mathcal{F}\left \{ \mathbf{x} \right \} \right\}\}
\end{equation}
For image and kernel sizes of $256 \times 256$ pixels, our implementation leads to nearly $500\times$ speedup: standard PyTorch convolution takes 2860ms, while Fourier convolution takes 5.92ms on a TITAN X. We also show how we can naturally extend our Fourier convolutions and FourierNet to a multiscale version using cropping in the Fourier domain in Appendix \ref{appendixmultiscale}.

\subsection{Physics-based autoencoder for simultaneous engineering of microscope encoder and reconstruction network decoder}
\label{autoencodertraining}
We describe the imaging process as the following transformation from the 3D light intensity volume $\mathbf{v}$ to the 2D image formed on the camera $\mathbf{c}$:
\begin{align}
  \bm{\upmu}_{\mathbf{c}} &= \mathbf{M}_{\bm{\upphi}}(\mathbf{v})\\
  \mathbf{c} &= \max{\left (\left [\bm{\upmu}_{\mathbf{c}} + \sqrt{\bm{\upmu}_{\mathbf{c}}}\epsilon\right ], 0 \right )}, \epsilon \sim \mathcal{N}(0, 1)
\end{align}
where $\mathbf{M}_{\bm{\upphi}}$ denotes the microscope parameterized by a 2D phase mask, $\bm{\upphi}$. This phase mask $\bm{\upphi}$ describes the 3D-to-2D encoding of this microscope model completely. A Poisson distribution with mean rate $\bm{\upmu}_{\mathbf{c}}$ describes the physics of photon detection at the camera, but sampling from this distribution is not differentiable. We approximate the noise distribution with a rectified Gaussian. We include details on $\mathbf{M}_{\bm{\upphi}}$ in Appendix \ref{appendixforwardsim} \cite{Goodman_2017}. Jointly training reconstruction networks and microscope parameters involves image simulation, reconstruction, then gradient backpropagation to update the reconstruction network and microscope parameters. Our parallelization strategy enables optimization of phase masks with millions of parameters in a feasible amount of time and memory per GPU. This also enables us to produce PSFs with multiple orders of magnitude more unique voxels than previous attempts (Appendix \ref{appendixpsftraining}). Details on parallelization, planewise reconstruction networks, and planewise sparse gradients are provided in Appendix \ref{appendixpsftraining}, Figure \ref{figsupparallelsim}.

We use the the normalized mean squared error (NMSE) as the basis for our loss function. Since we care more about the high spatial frequency content of the image which contains mostly cells, and less about the low spatial frequencies which contain mostly background, we implemented a two part loss function. $L_{\mathrm{HNMSE}}$ measures the NMSE between the high pass filtered volume and high pass filtered reconstruction. This is combined with the unfiltered NMSE $L_{\mathrm{NMSE}}$ between the original volume and its reconstruction. Formally, our loss function $L(\mathbf{v}, \mathbf{\hat{v}})$ for all snapshot microscopy reconstruction problems is defined as:
\begin{align}
  L(\mathbf{v}, \mathbf{\hat{v}}) &= L_{\mathrm{HNMSE}}(\mathbf{v}, \mathbf{\hat{v}}) + \beta L_{\mathrm{NMSE}}(\mathbf{v}, \mathbf{\hat{v}})\\
  L_{\mathrm{HNMSE}}(\mathbf{v}, \mathbf{\hat{v}}) &= \frac{\mathop{\mathbb{E}}\left [(H(\mathbf{v}) - H(\mathbf{\hat{v}}))^2\right ]}{\mathop{\mathbb{E}}(H(\mathbf{v})^2)}, L_{\mathrm{NMSE}}(\mathbf{v}, \mathbf{\hat{v}}) = \frac{\mathop{\mathbb{E}}\left [(\mathbf{v} - \mathbf{\hat{v}})^2\right ]}{\mathop{\mathbb{E}}(\mathbf{v}^2)}
\end{align}
where $H(\cdot)$ denotes high pass filtering and $\mathop{\mathbb{E}}(\cdot)$ denotes the mean over pixels and sample volumes. Both loss terms are normalized as shown to reduce variance in $L$, which can otherwise cause large magnitude fluctuations based on the brightness variation across training volumes and cause training instability. For our experiments, we set the weight $\beta$ for the $L_{\mathrm{NMSE}}$ term to 0.1.
\titlespacing{\section}{0pt}{2ex}{1ex}
\titlespacing{\subsection}{0pt}{1ex}{1ex}

\section{Results}
\label{results}

\textbf{DiffuserCam Lensless Mirflickr Dataset}
We also test reconstruction performance on experimental computational photography data\footnote{Publicly available: \url{https://waller-lab.github.io/LenslessLearning/dataset.html}} from \cite{Monakhova_Yurtsever_Kuo_Antipa_Yanny_Waller_2019} (Figure \ref{figdlmdcomparison}). This is a dataset constructed by displaying RGB color natural images from the MIRFlickr dataset on a monitor and then capturing diffused images by the DiffuserCam lensless camera. The dataset contains 24,000 pairs of DiffuserCam and ground truth images. The goal of the dataset is to learn to reconstruct the ground truth images from the diffused images. As in \cite{Monakhova_Yurtsever_Kuo_Antipa_Yanny_Waller_2019}, we train on 23,000 paired diffused and ground truth images, and test on 999 held-out pairs of images.

\textbf{Larval Zebrafish Snapshot Microscopy Dataset}
We show all our results for snapshot microscopy using our simulation of a snapshot microscope, and our engineered optical encoders have not been experimentally tested for reconstruction performance on a programmable microscope. We simulate snapshot imaging using high resolution confocal imaging volumes of zebrafish. These are volumes of transgenic larval zebrafish whole brains expressing nuclear-restricted GCaMP6 calcium indicator in all neurons. These images are representative of brain-wide activity imaging. We train on 58 different zebrafish volumes (which we augment heavily) and test on 10 held-out volumes. For all experiments, we downsample the high resolution confocal data to (1.0 $\upmu$m z, 1.625 $\upmu$m y, 1.625 $\upmu$m x). We created 4 datasets from these scanned volumes corresponding to imaging different fields of view or regions of interest (ROIs) for our experiments. Full specifications for these datasets are in Table \ref{tabdatasets} (Appendix \ref{appendixpsftraining}). For Figure \ref{figresultscomparison}, we restrict the field of view to (200 $\upmu$m z, 416 $\upmu$m y, 416 $\upmu$m x) with a tall cylinder cutout of diameter 193 $\upmu$m and height 200 $\upmu$m and image with $256 \times 256$ pixels on the simulated camera sensor. Figure \ref{figaperturecomparison} and Table \ref{tabmatrix} show our larger experiments with $512 \times 512$ simulated camera pixels, with a field of view of (250 $\upmu$m z, 832 $\upmu$m y, 832 $\upmu$m x).

\subsection{FourierNets outperform state-of-the-art for reconstructing natural images captured by DiffuserCam lensless camera}
\begin{figure}[h]
  \includegraphics[width=\linewidth]{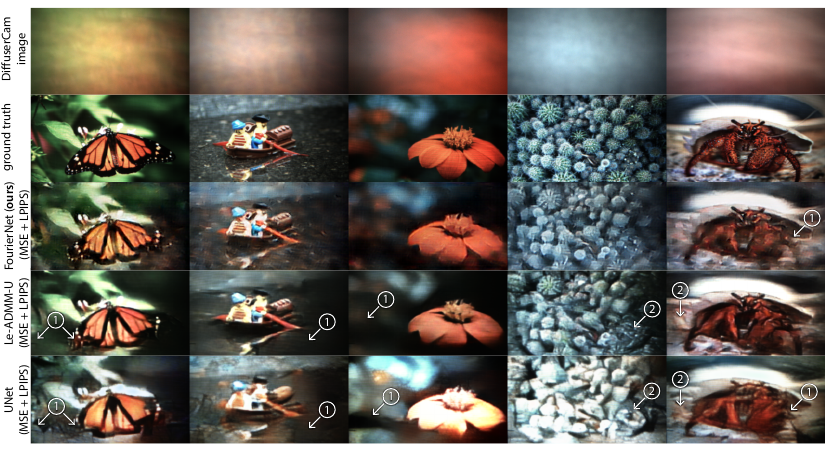}
  \caption{Comparisons of our method (third row) to state-of-the-art learned reconstruction methods on lensless diffused images of natural scenes. Regions labeled {\Large \textcircled{\normalsize 1}} indicate missing details, either resolution or textures in backgrounds. Regions labeled {\Large \textcircled{\normalsize 2}} indicate hallucinated textures. Note that the previous state-of-the-art solution (fourth row) \cite{Monakhova_Yurtsever_Kuo_Antipa_Yanny_Waller_2019} exhibits both issues more often compared to our method.}
  \label{figdlmdcomparison}
\end{figure}

\begin{table}[h]
\centering
    \caption{Quality of natural image reconstruction on the DiffuserCam Lensless Mirflickr Dataset (mean \(\pm\) s.e.m., \(n = 999\)). Superscripts denote loss function: \textsuperscript{1} MSE, \textsuperscript{2} MSE+LPIPS.}
\label{tabdlmd}
\begin{tabular}{@{}p{27.8mm}@{\hskip 1.0mm}p{23.1mm}@{\hskip 1.0mm}p{17mm}lp{17.5mm}p{17.5mm}@{}}
    \toprule
    \textbf{Method} & \textbf{MSE} \(\downarrow\) (\(\times10^{-2}\)) & \textbf{LPIPS} \(\downarrow\) & \textbf{MS-SSIM} \(\uparrow\) & \textbf{PSNR} \(\uparrow\) & \textbf{Time} \(\downarrow\) (ms)\\
    \midrule
    FourierNet$^1$ & \textbf{0.39} $\pm$ \textbf{0.007} & 0.20 $\pm$ 0.00 & \textbf{0.882} $\pm$ \textbf{0.001} & \textbf{24.8} $\pm$ \textbf{0.09} & 35.54\\
    FourierNet$^2$ & 0.54 $\pm$ 0.010 & \textbf{0.16} $\pm$ \textbf{0.00} & 0.868 $\pm$ 0.001 & 23.4 $\pm$ 0.09 & 35.54\\
    \midrule
    Le-ADMM-U$^2$ \cite{Monakhova_Yurtsever_Kuo_Antipa_Yanny_Waller_2019} & 0.75 $\pm$ 0.021 & 0.19 $\pm$ 0.00 & 0.865 $\pm$ 0.002 & 22.1 $\pm$ 0.09 & 48.59\\
    UNet$^2$ \cite{Monakhova_Yurtsever_Kuo_Antipa_Yanny_Waller_2019} & 1.68 $\pm$ 0.060 & 0.24 $\pm$ 0.00 & 0.818 $\pm$ 0.002 & 19.2 $\pm$ 0.11 & \textbf{06.97}\\
    \bottomrule
\end{tabular}
\end{table}

We compare our FourierNet architecture to the best learned method from \cite{Monakhova_Yurtsever_Kuo_Antipa_Yanny_Waller_2019} using an unrolled ADMM and a denoising UNet, as well as to a vanilla UNet from \cite{Monakhova_Yurtsever_Kuo_Antipa_Yanny_Waller_2019}. Architecture details are in Appendix \ref{appendiximplementationdetails}, \ref{appendixdlmd}. We can see that FourierNet visually outperforms the methods from \cite{Monakhova_Yurtsever_Kuo_Antipa_Yanny_Waller_2019} in Figure \ref{figdlmdcomparison}, and quantitatively in Table \ref{tabdlmd} across all quality metrics. The Le-ADMM-U and UNet results in Table \ref{tabdlmd} were reported by \cite{Monakhova_Yurtsever_Kuo_Antipa_Yanny_Waller_2019} using a combined MSE + LPIPS loss. Unlike \cite{Monakhova_Yurtsever_Kuo_Antipa_Yanny_Waller_2019}, we find that training FourierNets with the MSE loss alone provides reconstructions visually similar to the ground truth as shown in Figure \ref{figsupdlmdcomparison} (Appendix \ref{appendixdlmd}). Timings in Table \ref{tabdlmd} are for only the forward pass on a single TITAN Xp GPU.

\subsection{FourierNets outperform UNets for engineering non-local optical encoders}
\label{microscopeoptimizationcomparison}

\begin{table}[h]
\centering
\caption{Quality of reconstructed volumes after optimizing microscope parameters to image zebrafish on $256 \times 256$ pixel camera (mean $\pm$ s.e.m., $n = 10$)}
\label{tabmetrics}
\begin{tabular}{@{}lllll@{\hskip 1.0mm}c@{}}
    \toprule
    \textbf{Microscope} & \textbf{Reconstruction} & $\bm{L_{\mathrm{HNMSE}}}$ $\downarrow$ & \textbf{MS-SSIM} $\uparrow$ & \textbf{PSNR} $\uparrow$ & \textbf{Time} $\downarrow$ (s)\\
    \midrule
    FourierNet2D & FourierNet3D & \textbf{0.6093} $\pm$ \textbf{0.0209} & \textbf{0.955} $\pm$ \textbf{0.004} & \textbf{34.89} $\pm$ \textbf{0.88} & \textbf{0.38}\\
    FourierNet2D & UNet3D & 0.7298 $\pm$ 0.0151 & 0.923 $\pm$ 0.008 & 30.16 $\pm$ 0.94 & 0.96\\
    Wiener + UNet & Wiener + UNet & 0.7223 $\pm$ 0.0179 & \textbf{0.957} $\pm$ \textbf{0.003} & \textbf{34.49} $\pm$ \textbf{0.91} & 0.73\\
    UNet2D & UNet3D & 0.7109 $\pm$ 0.0161 & 0.913 $\pm$ 0.009 & 29.17 $\pm$ 1.13 & 0.96\\
    \bottomrule
\end{tabular}
\end{table}

\begin{figure}[h]
  \includegraphics[width=\linewidth]{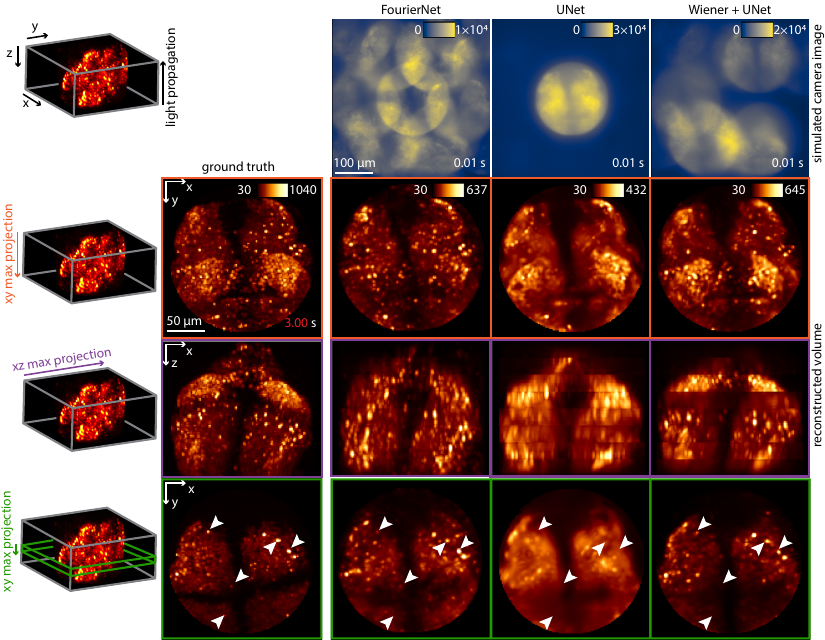}
  \caption{Comparing simulated camera images (0.01 second expected acquisition time) and corresponding reconstructions of a volume captured using our FourierNet (left) versus UNet (middle) and Wiener + UNet \cite{Ikoma_Nguyen_Metzler_Peng_Wetzstein_2021} (right) optimized microscopes. Top row shows simulated $256\times256$ pixel camera images; bottom right of camera image shows approximate acquisition time (given a reasonable number of simulated photons per pixel, i.e. SNR). Ground truth has no corresponding camera image, because the 3D volume is imaged directly via slow high resolution confocal microscopy (3 second acquisition time). Colored arrows in left column show projection axis for each row. White arrows show individual neurons clearly visible for FourierNet, but not for UNet. Wiener + UNet \cite{Ikoma_Nguyen_Metzler_Peng_Wetzstein_2021} has detail in some planes, but not consistently throughout the whole volume, and also uses fewer camera pixels.}
  \vspace{-0.6cm}
  \label{figresultscomparison}
\end{figure}
We compare optimizing microscope parameters $\bm{\upphi}$ with three neural networks: 1) using our FourierNet with 2D convolutions (FourierNet2D), 2) using a vanilla UNet with 2D convolutions (UNet2D), and 3) using a Wiener deconvolution as an approximate inverse combined with a refining UNet, as in \cite{Ikoma_Nguyen_Metzler_Peng_Wetzstein_2021}. Training is a two stage process in which a plane-wise reconstruction network with fewer parameters and 2D convolutions is used during optimization of $\bm{\upphi}$, then once $\bm{\upphi}$ is fixed a more powerful reconstruction network with 3D convolutions is used, except for the Wiener + UNet \cite{Ikoma_Nguyen_Metzler_Peng_Wetzstein_2021} method which is trained in a single stage as in \cite{Ikoma_Nguyen_Metzler_Peng_Wetzstein_2021}.  Table \ref{tabmetrics} shows the type of network used for the first stage to train the optical encoder $\mathbf{M}_{\bm{\upphi}}$ in the first column, and the type of network used in the second stage for training the reconstruction network only (given a fixed $\bm{\upphi}$) in the second column. We find this scheme achieves better reconstruction quality because the reconstruction network does not need to constantly adapt to a changing optical encoding (Appendix \ref{appendiximplementationdetails}). We train these microscopes and reconstruction networks on ROIs which are tall cylindrical cutouts of zebrafish with diameter 193 $\upmu$m and height 200 $\upmu$m. Sample volumes are imaged on a camera with $256 \times 256$ pixels (Figure \ref{figresultscomparison}). FourierNet2D has 2 convolution layers (with 99.8\% of its kernel parameters in the initial Fourier convolution layer), while UNet2D has 32 convolution layers (kernel parameters approximately uniformly distributed per layer). UNet2D was designed to have a global receptive field. FourierNet2D and UNet2D both have $\sim 4 \times 10^7$ parameters; FourierNet3D has $\sim 6 \times 10^7$ parameters vs. $\sim 10^8$ for UNet3D. The Wiener + UNet method \cite{Ikoma_Nguyen_Metzler_Peng_Wetzstein_2021} has $\sim 8 \times 10^7$ parameters. Architecture details are in Appendix \ref{appendiximplementationdetails}, \ref{appendixnetworkcomparison}.

Simulated camera images (Figure \ref{figresultscomparison}) show that the UNet microscope does not make sufficient use of camera pixels, producing only a single view of the volume. We speculate this is due to a local information prior in the small kernels of UNets.  Adding a Wiener deconvolution as an approximate inverse before a UNet also does not result in a microscope that makes full use of camera pixels, though better than the UNet alone. The FourierNet microscope uses more camera pixels and performs better than the UNet microscope for reconstruction (Figure \ref{figresultscomparison}, quantified in Table \ref{tabmetrics}). Timings in Table \ref{tabmetrics} are for only the forward pass on a single TITAN Xp GPU; one training iteration on 8 GPUs takes $\sim$0.4 seconds for FourierNet3D, $\sim$0.8 seconds for Wiener + UNet, and $\sim$0.8 seconds for UNet3D (Appendix \ref{appendiximplementationdetails}). Both reconstruction networks must reconstruct from images that have a compressed encoding of 3D information, but the FourierNet2D is clearly more effective than the UNet2D at optimizing this encoding.

\subsection{FourierNets outperform UNets for 3D snapshot microscopy volume reconstruction}
\label{networkcomparison}

\begin{table}
\centering
\caption{ROI specific microscope parameter optimization for 3 types of zebrafish volumes (mean PSNR (top), MS-SSIM (bottom) $\pm$ s.e.m., $n = 10$). Green shows regions of interest.}
\label{tabmatrix}
\begin{tabular}{@{}llll@{}}
    \toprule
    ~ & \multicolumn{3}{c}{\textbf{Microscope parameters optimized for}} \\
    ~ & Type A & Type B & Type C\\
    \textbf{Tested on} & \includegraphics[scale=0.5]{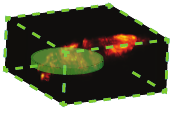} & \includegraphics[scale=0.5]{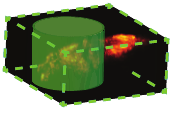} & \includegraphics[scale=0.5]{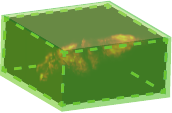}\\
    \midrule
    Type A & \parbox[t]{2cm}{\textbf{49.76} $\pm$ \textbf{1.35}\\\textbf{0.998} $\pm$ \textbf{0.000}} & \parbox[t]{2cm}{46.03 $\pm$ 1.33\\0.996 $\pm$ 0.001} & \parbox[t]{2cm}{42.67 $\pm$ 1.14\\0.992 $\pm$ 0.002}\\\midrule
    Type B & \parbox[t]{2cm}{35.56 $\pm$ 1.41\\0.965 $\pm$ 0.004} & \parbox[t]{2cm}{\textbf{37.34} $\pm$ \textbf{0.96}\\\textbf{0.972} $\pm$ \textbf{0.003}} & \parbox[t]{2cm}{35.38 $\pm$ 1.16\\0.967 $\pm$ 0.003}\\\midrule
    Type C & \parbox[t]{2cm}{30.87 $\pm$ 1.15\\0.912 $\pm$ 0.007} & \parbox[t]{2cm}{31.48 $\pm$ 0.93\\0.920 $\pm$ 0.006} & \parbox[t]{2cm}{\textbf{33.79} $\pm$ \textbf{0.90}\\\textbf{0.935} $\pm$ \textbf{0.006}}\\
    \bottomrule
\end{tabular}
\vspace{-0.3cm}
\end{table}

We can determine which architecture is better for volume reconstruction by choosing fixed microscope parameters and varying the architecture, except for the Wiener + UNet method \cite{Ikoma_Nguyen_Metzler_Peng_Wetzstein_2021} which is trained in one stage. In Table \ref{tabmetrics}, we compare results using a FourierNet with 3D convolutions (FourierNet3D) and a vanilla UNet with 3D convolutions (UNet3D). UNet3D was also designed to have a global receptive field. Architecture details are in Appendix \ref{appendiximplementationdetails}, \ref{appendixnetworkcomparison}.

Reconstruction results in Table \ref{tabmetrics} compare normalized MSE $L_{\mathrm{HNMSE}}$ between the high pass filtered volume and high pass filtered reconstruction, the multiscale structural similarity $\mathrm{MS-SSIM}$ between the true volume and its reconstruction, and finally the peak signal-to-noise ratio $\mathrm{PSNR}$. We also visualize reconstruction results for a volume in the head of a zebrafish in Figure \ref{figresultscomparison}. The UNet3D reconstruction networks (using either microscope) fall significantly behind the FourierNet3D reconstruction network in all metrics, despite their global receptive field. The Wiener + UNet \cite{Ikoma_Nguyen_Metzler_Peng_Wetzstein_2021} network achieves similar $\mathrm{MS-SSIM}$ and $\mathrm{PSNR}$ as the FourierNet but a worse $L_{\mathrm{HNMSE}}$ due to the inconsistent detail, though the reconstructions are slightly better than the UNet3D for certain regions of the volume.
\vspace{-0.1in}
\subsection{Engineered optical encoders can be optimized for a region of interest}
\label{aperturecomparison}

\begin{figure}
  \includegraphics[width=\linewidth]{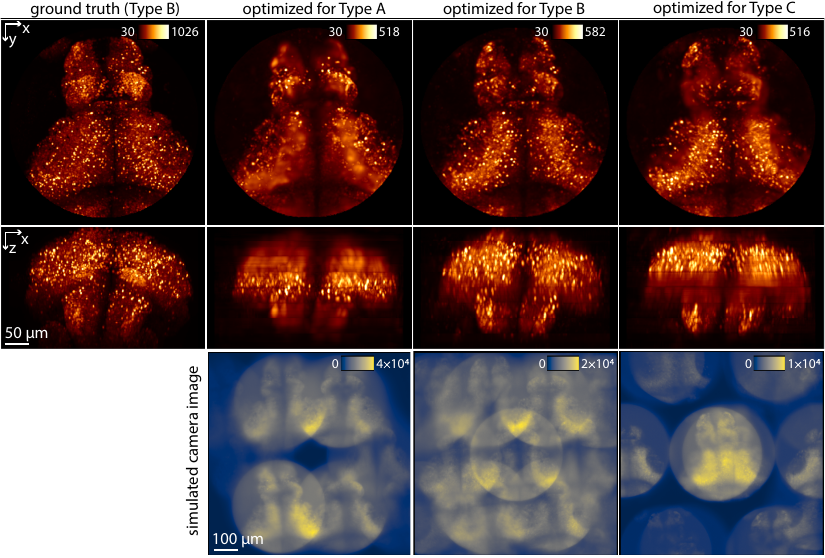}
  \caption{Reconstructed volumes resulting from imaging Type B ROIs by microscopes optimized for Types A, B, and C. Imaging Type B ROIs with microscope parameters optimized for Type B ROIs yields the best reconstructions. Top to bottom: xy max projection, xz max projection, simulated $512\times512$ camera image.}
  \label{figaperturecomparison}
  \vspace{-0.5cm}
\end{figure}

To explore the effect of ROI size on optimized $\bm{\upphi}$ and the resulting reconstruction performance, we optimized $\bm{\upphi}$s for three different regions of interest: 1) Type A, ROIs with a short cylinder cutout of 386 $\upmu$m diameter and 25 $\upmu$m height, 2) Type B, ROIs with a tall cylinder cutout of 386 $\upmu$m diameter and 250 $\upmu$m height, and 3) Type C, ROIs without any cutout of dimension (250 $\upmu$m z $\times$ 832 $\upmu$m y $\times$ 832 $\upmu$m x) (Table \ref{tabmatrix}). All ROI types were imaged with $512 \times 512$ pixels on the simulated camera. We then tested reconstruction performance on all combinations of optimized $\bm{\upphi}$ and ROI type, as shown in Table \ref{tabmatrix} and visualized in Figure \ref{figaperturecomparison} for Type B. We include architecture details in Appendix \ref{appendiximplementationdetails}, \ref{appendixaperturecomparison}.

We see in Table \ref{tabmatrix} that for all types, highest performance is achieved using the phase mask optimized for that particular type. These results show that there is potentially a large benefit to optimizing type-specific optical encoders, which can be easily implemented in a programmable microscope.

\subsection{Engineered optical encoders can be implemented on a programmable microscope}
We demonstrate that the optical encoders engineered by simulations can be implemented on real hardware using a prototype programmable microscope. Our engineered phase masks were displayed on a spatial light modulator (SLM) located in the pupil plane. We observe a good qualitative match between the features of the simulated and experimental optical encoders (point spread functions) in Figure \ref{figmeasuredpsf}, technical details in Appendix \ref{appendixmeasurement}. There is also a good qualitative match between the simulated camera images generated by both optical encoders. These results demonstrate the potential for programmable microscopy with spatial light modulators.

We also note that our optimized optical encoders result in pencil-like elements, qualitatively similar to lenslet-based approaches \cite{Yanny_Antipa_Liberti_Dehaeck_Monakhova_Liu_Shen_Ng_Waller_2020, Linda_Liu_Kuo_Antipa_Yanny_Waller_2020, Cong_Wang_Chai_Hang_Shang_Yang_Bai_Du_Wang_Wen_2017}. However as we show in our supplement, our optimized phase masks (Figures \ref{figsuptypeabcphase}) appear qualitatively different from lenslet-based phase masks, with different regions of the pupil contributing to the same pencil. This suggests a qualitatively different, and perhaps more light efficient, mechanism for generating high resolution projections of the volume along different axes.

\begin{figure}[h]
  \includegraphics{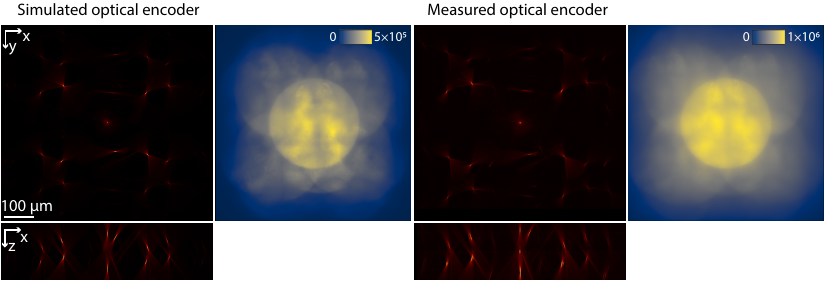}
  \caption{Simulated versus measured optical encoders (point spread functions) with corresponding simulated camera images using an example from our Type B zebrafish dataset. Zoom in for best viewing.}
  \label{figmeasuredpsf}
  \vspace{-0.4cm}
\end{figure}

\section{Discussion}
\label{discussion}
\vspace{-0.3cm}
\textbf{Summary}
We have presented the FourierNet architecture as a deep network based decoder for computational imaging with highly non-local optical encoders. We demonstrated the superiority of FourierNets for lensless photography on the DiffuserCam dataset, and also for the end-to-end optimization of highly non-local optical encodings (where UNets fail) for 3D snapshot microscopy. FourierNets are many orders of magnitude faster than traditional iterative reconstruction algorithms \cite{Monakhova_Yurtsever_Kuo_Antipa_Yanny_Waller_2019} and generate higher quality reconstructions by learning image priors. Generally, our global Fourier-domain convolution architecture could be applicable to other problems where global integration of features is necessary, though we have focused on computational imaging applications where physics-based optical encoders induce global mixing of information. Our work solves two important computational problems preventing E2E optimization of large highly non-local optical encoders: the computational complexity of simulating large non-local encoders, and the effective decoding from such encoders. Our contributions are primarily computational, however we also demonstrate the potential for implementing 3D snapshot microscopy with an E2E optimized non-local optical encoder using an SLM-based programmable microscope. And in simulation, we demonstrate the potential for ROI-specific optimization of optical encoders for 3D snapshot microscopy, which can be efficiently implemented in a programmable microscope.

\textbf{Limitations}
While we have fully demonstrated that our framework now enables computational E2E optimization of large non-local optical encoders, we do not focus on their hardware implementation. The specific claims regarding the potential benefits of ROI-specific optimization of optical encoders are only evaluated in simulation, and have only been evaluated for whole brain larval zebrafish data.

\textbf{Reproducibility}
We train on 8 Quadro RTX 8000 GPUs for the largest experiments, and have described our pre-processing, training, and testing procedures in Appendix \ref{appendixpsftraining}, \ref{appendiximplementationdetails}, \ref{appendixnetworkcomparison}. We have made our simulation software, training scripts, and our datasets available at \url{https://github.com/TuragaLab/snapshotscope}.

\textbf{Broader Impact}
End-to-end optimization of optics can improve image quality but also enable multiplexed imaging not currently possible with standard optics. Our PyTorch-based optical modeling library as well as our neural network architectures are publicly available and could enable new experiments in neuroscience via whole-brain imaging at an order of magnitude greater temporal resolution. Our training procedure does require long optimization periods with many GPUs, which poses a carbon footprint and barrier to usage compared to conventional microscopy.

\section*{Acknowledgement}
We would like to thank William Bishop, Roman Vaxenburg, Gert-Jan Both, Janne Lappalainen, Nathan Klapoetke, Richard Xu, Lu Mi, Sridhama Prakhya, and Jinyao Yan for invaluable feedback and discussions. We thank Howard Hughes Medical Institute and a Simons Foundation grant (Simons Collaboration on the Global Brain, 542943SPI, Ahrens) for their funding. We also thank Huazhong University of Science and Technology and the China Scholarship Council for supporting the work of Zhenfei Jiao. We additionally thank the Janelia Visiting Scientist Program for supporting the work of Ruth Sims.

\medskip

{
\bibliographystyle{ieeetr}
\bibliography{references}

}

\section*{Checklist}


\begin{enumerate}

\item For all authors...
\begin{enumerate}
  \item Do the main claims made in the abstract and introduction accurately reflect the paper's contributions and scope?
    \answerYes{See our summary in Section \ref{discussion}.}
  \item Did you describe the limitations of your work?
    \answerYes{We also discuss limitations in Section \ref{discussion}.}
  \item Did you discuss any potential negative societal impacts of your work?
    \answerYes{We also discuss broader impacts in Section \ref{discussion}.}
  \item Have you read the ethics review guidelines and ensured that your paper conforms to them?
    \answerYes{}
\end{enumerate}

\item If you are including theoretical results...
\begin{enumerate}
  \item Did you state the full set of assumptions of all theoretical results?
    \answerNA{}
  \item Did you include complete proofs of all theoretical results?
    \answerNA{}
\end{enumerate}

\item If you ran experiments...
\begin{enumerate}
  \item Did you include the code, data, and instructions needed to reproduce the main experimental results (either in the supplemental material or as a URL)?
    \answerYes{See Appendix \ref{appendixpsftraining}, \ref{appendiximplementationdetails}, \ref{appendixnetworkcomparison}. We have also released our simulation and optimization library (along with instructions to download data) at \url{https://github.com/TuragaLab/snapshotscope}.}
  \item Did you specify all the training details (e.g., data splits, hyperparameters, how they were chosen)?
    \answerYes{See Appendix \ref{appendixpsftraining}, \ref{appendiximplementationdetails}, \ref{appendixnetworkcomparison}.}
  \item Did you report error bars (e.g., with respect to the random seed after running experiments multiple times)?
    \answerYes{See standard error of the mean values in Table \ref{tabmetrics}.}
  \item Did you include the total amount of compute and the type of resources used (e.g., type of GPUs, internal cluster, or cloud provider)?
    \answerYes{See limitations in Section \ref{discussion} and Appendix \ref{appendiximplementationdetails}.}
\end{enumerate}

\item If you are using existing assets (e.g., code, data, models) or curating/releasing new assets...
\begin{enumerate}
  \item If your work uses existing assets, did you cite the creators?
    \answerYes{We have cited UNets as \cite{Ronneberger_Fischer_Brox_2015} and DLMD as \cite{Monakhova_Yurtsever_Kuo_Antipa_Yanny_Waller_2019}.}
  \item Did you mention the license of the assets?
    \answerNA{}
  \item Did you include any new assets either in the supplemental material or as a URL?
    \answerYes{We have released our simulation and reconstruction network code at \url{https://github.com/TuragaLab/snapshotscope}.}
  \item Did you discuss whether and how consent was obtained from people whose data you're using/curating?
     \answerYes{See footnote 1 about DiffuserCam dataset. Zebrafish data is our own.}
  \item Did you discuss whether the data you are using/curating contains personally identifiable information or offensive content?
    \answerNA{}
\end{enumerate}

\item If you used crowdsourcing or conducted research with human subjects...
\begin{enumerate}
  \item Did you include the full text of instructions given to participants and screenshots, if applicable?
    \answerNA{}
  \item Did you describe any potential participant risks, with links to Institutional Review Board (IRB) approvals, if applicable?
    \answerNA{}
  \item Did you include the estimated hourly wage paid to participants and the total amount spent on participant compensation?
    \answerNA{}
\end{enumerate}

\end{enumerate}


\newpage
\appendix
\section{Appendix}
\subsection{Forward simulation of programmable 3D snapshot microscope}
\label{appendixforwardsim}
\begin{figure}[!h]
  \includegraphics[width=\linewidth]{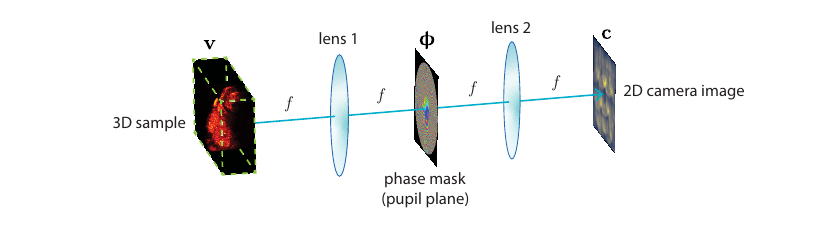}
  \caption{Diagram of a $4f$ optical model that is the basis for our simulated microscope $\mathbf{M}_{\bm{\upphi}}$, showing the Fourier plane in which we have the programmable and trainable 2D phase mask $\bm{\upphi}$.}
  \label{figsup4fmodel}
\end{figure}

Here we describe our wave optics simulation of the microscope $\mathbf{M}_{\bm{\upphi}}$, which we model as a $4f$ system \cite{Goodman_2017}. The $4f$ optical system consists of two lenses, the first spaced one focal length from the object plane and the second spaced one focal length away from one focal length beyond the first lens (Figure \ref{figsup4fmodel}). In between these two lenses, we can place a phase mask to manipulate the light field before passing through the second lens and forming an image on the camera sensor.

We are concerned here with fluorescence microscopy, meaning that the sources of light that we image are individual fluorescent molecules, which we can model as point emitters. Because these molecules emit incoherent light, the camera sensor in effect sums the contributions of each point emitter. In order to model such an imaging system, we first need to address modeling a single point emitter's image on the camera.

We can analytically calculate the complex-valued light field one focal length after the first lens (which we call the pupil plane) due to a point source centered at some plane $z$ (where $z$ is a distance from the object plane $z = 0$). If the point source were centered ($x=0, y=0$) in the object focal plane $z = 0$, we would have a plane wave at the pupil plane, but for the more general case of a point source at an arbitrary plane $z$ relative to the object plane $z = 0$, we can analytically calculate the complex-valued light field entering the pupil plane:
\begin{equation}
    u_{\mathrm{point}}(\mathbf{k}; z) = \exp\left[i 2\pi z\sqrt{\left(\frac{n}{\lambda}\right)^2 - \left\vert\left\vert\mathbf{k}\right\vert\right\vert_2^2}\right]
\end{equation}
where $u_{\mathrm{point}}$ is the incoming light field entering the pupil due to a point source centered in the plane at $z$, $\mathbf{k} \in \mathbb{R}^2$ denotes frequency space coordinates of the light field in the pupil plane, $n$ is the refractive index, and $\lambda$ is the wavelength of light \cite{Goodman_2017}.

In this pupil plane, we can then apply a phase mask $\bm{\upphi}$ to the light field, which is modeled as a multiplication of $u_{\mathrm{point}}(\mathbf{k}; z)$ and $e^{i\bm{\upphi}(\mathbf{k})}$, the complex phase of the pupil function. The light field exiting the pupil is therefore described by
\begin{equation}
    u_{\mathrm{pupil}}(\mathbf{k}; z) = u_{\mathrm{point}} (\mathbf{k}; z) p(\mathbf{k})
\end{equation}
where $p(\mathbf{k})$ is the pupil function, composed of an amplitude $a(\mathbf{k})$ and phase $\bm{\upphi}(\mathbf{k})$:
\begin{align}
    p(\mathbf{k}) &= a(\mathbf{k}) e^{i\bm{\upphi}(\bf{k})}\\
    a(\mathbf{k}) &=
        \begin{cases}
        1 & \left\vert\left\vert\mathbf{k}\right\vert\right\vert_2 \leq \frac{\mathrm{NA}}{\lambda}\\
        0 & \left\vert\left\vert\mathbf{k}\right\vert\right\vert_2 > \frac{\mathrm{NA}}{\lambda}
        \end{cases}
\end{align}
where $\mathrm{NA}$ is the numerical aperture of the lens \cite{Goodman_2017}.

The light field at the camera plane can then be described by a Fourier transform \cite{Goodman_2017}:
\begin{equation}
    \bf{u}_{\mathrm{camera}} = \mathcal{F}\left\{\bf{u}_{\mathrm{pupil}}\right\}
\end{equation}

The camera measures the intensity of this complex field:
\begin{equation}
    \label{eqpsf}
    s(\mathbf{x}; z) = \left\vert u_{\mathrm{camera}}(\mathbf{x}; z) \right\vert^2
\end{equation}
where $\mathbf{x} \in \mathbb{R}^2$ denotes spatial coordinates in the camera plane \cite{Goodman_2017}.

We can call this intensity $\mathbf{s}$ the point response function (PRF). If the shape of the PRF is translationally equivariant in $\mathbf{x}$, meaning that moving a point source in-plane creates the same field at the camera, just shifted by the corresponding amount, then we call this PRF a point spread function (PSF), which we also refer to as an optical encoder. Note that moving the point source in $z$ will not give the same shape, which allows our system to encode depth information through the PSF \cite{Broxton_2017}.

In order to avoid edge effects during imaging, we simulate the PSF at a larger field of view, then crop and taper the edges of the PSF:
\begin{equation}
    \bf{s}_{\mathrm{taper}} = \mathrm{crop}[\bf{s}] \odot \bf{t}
\end{equation}
where $\bf{t}$ is a taper function created by taking the sigmoid of a distance transform divided by a width factor controlling how quickly the taper goes to 0 at the edges and $\odot$ denotes elementwise multiplication. We intentionally simulate a larger field of view than the sample volume in order to avoid edge artifacts. The purpose of the $\mathrm{crop}[\cdot]$ is to cut the PSF to the correct field of view. The purpose of the tapering is to remove artifacts at the edges of the cropped PSF. After we compute this cropped and tapered PSF, we also downsample $\mathbf{s}_{\mathrm{taper}}$ to the size of the data $\mathbf{v}$ in order to save memory.

Imaging is equivalent to the convolution of the incoming light field volume intensity $\mathbf{v}$ and the cropped and tapered PSF $\mathbf{s}_{\mathrm{taper}}$ for a given plane. At the camera plane, the light field intensity is measured by the camera sensor. Therefore, we can describe the forward model as the following convolution and integral over planes:
\begin{equation}
  \label{eqimaging}
  \bm{\upmu}_{\mathbf{c}}(\mathbf{x}) = \iint \! v(\bm{\tau}_{\mathbf{x}}; z) s_{\mathrm{taper}}(\mathbf{x} - \bm{\tau}_{\mathbf{x}}; z)\, \mathrm{d\bm{\tau}_{\mathbf{x}}}\, \mathrm{d}z
\end{equation}

We then model shot noise of the camera sensor to produce the final image $\mathbf{c}$, for which the appropriate model is sampling from a Poisson distribution with a mean of $\bm{\upmu}_{\mathbf{c}}$ \cite{Goodman_2017}:
\begin{equation}
  \mathbf{c} \sim \mathrm{Poisson}\left( \bm{\upmu}_{\mathbf{c}} \right)\\
\end{equation}
However, because we cannot use the reparameterization trick to take pathwise derivatives through the discrete Poisson distribution, we instead approximate the noise model with a rectified Gaussian distribution:
\begin{align}\
  \epsilon &\sim \mathcal{N}(0, 1)\\
  \mathbf{c} &\approx \max{\left (\left [\bm{\upmu}_{\mathbf{c}} + \sqrt{\bm{\upmu}_{\mathbf{c}}}\epsilon\right ], 0 \right )}
\end{align}

We now turn our attention to selecting the number of pixels used in the phase mask, i.e. the number of parameters for $\mathbf{M}_{\bm{\upphi}}$. We first need to determine the pixel size for Nyquist sampling the image plane with an objective of a given NA (numerical aperture). For a given pixel size $\Delta x$, we know that in frequency space coordinates we will have a bandwidth of $\frac{1}{\Delta x}$, spanning $-\frac{1}{2\Delta x}$ to $\frac{1}{2\Delta x}$. Because we must have
\begin{align}
    \left\vert\left\vert\mathbf{k}\right\vert\right\vert_2 \leq \frac{\mathrm{NA}}{\lambda}
\end{align}
we know that the Nyquist sampling pixel size is given by
\begin{align}
    \Delta x^* = \frac{\lambda}{2\mathrm{NA}}.
\end{align}

Therefore, in the image plane, for a desired field of view $L$ we must have at least
\begin{align}
    N^* = \frac{L}{\Delta x^*}
\end{align} pixels. The discretization in the pupil plane will be the same, which means we will need to have at least $N^*$ pixels in the pupil plane to achieve the appropriate light field in the image plane. For our settings of NA = 0.8, $\lambda = 0.532 \upmu\mathrm{m}$, and $L = 823 \upmu\mathrm{m}$, we have $\Delta x^* = 0.3325 \upmu\mathrm{m}$ and $N^* = 2476$ pixels. Thus, a reasonable choice is $\Delta x = 0.325 \upmu\mathrm{m}$ and $N = 2560$ pixels. Note that these simulation parameters are independent of the camera pixels; we have determined only how many pixels must be used in the phase mask in order to ensure our PSF can occupy the full field of view. The camera sensor can sample the field at the image plane at an independent pixel size.

\subsection{Multiscale feature extraction using FourierUNets}
\label{appendixmultiscale}
\begin{figure}[h]
  \includegraphics[width=\linewidth]{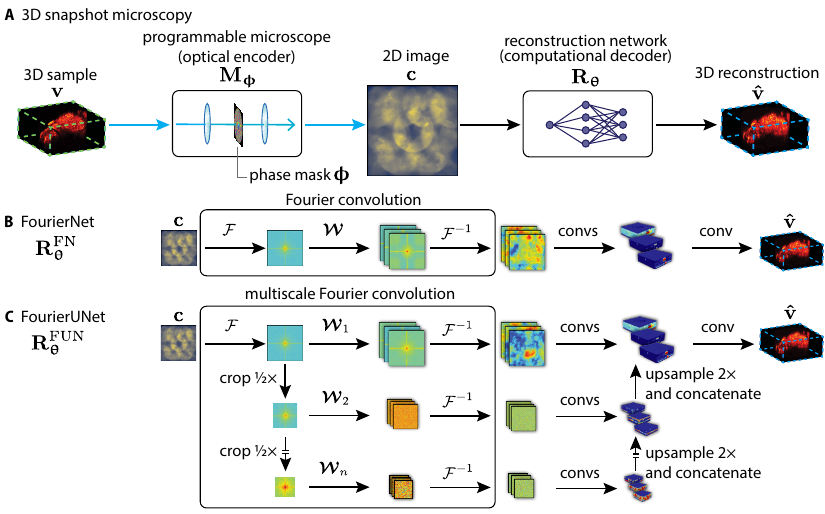}
  \caption{Overview of our problem setup and our proposed network architectures. Top row (\textbf{A}) shows the problem of 3D snapshot microscopy, where we computationally reconstruct a 3D volume from a 2D image. Middle row (\textbf{B}) shows our proposed FourierNet architecture, which includes a \textbf{Fourier convolution} layer that enables efficient computation of global features. Bottom row (\textbf{C}) shows an extension of our proposed architecture, the FourierUNet, which mimics the multiscale feature extraction of a standard UNet efficiently and with global features using a \textbf{multiscale Fourier convolution}.}
  \label{figsupoverview}
\end{figure}

\textbf{FourierUNet} We propose a multi-scale variant of the FourierNet by bringing together elements of the multi-scale UNet and the single-scale FourierNet. Here, we take advantage of the fact that down-sampling in image space corresponds to a simple cropping operation in the Fourier domain, resulting in a band-limited computation of a feature map. We efficiently implement multi-scale global Fourier convolutions (Figure \ref{figoverview}C) to replace the encoding/``analysis'' pathway of a UNet. We then use the standard decoding/``synthesis'' pathway of the UNet to combine the multi-scale features into a single 3D volume reconstruction (Appendix \ref{appendiximplementationdetails}, \ref{appendixnetworkcomparison}). Thus we can study whether multi-scale features or global context is more important for decoding non-local optical encoders.

\textbf{Multi-scale Fourier domain convolutions}
It is well-known \cite{Oppenheim_Willsky_2013} that downsampling corresponds to cropping in the Fourier domain. Thus the Fourier convolution can be extended to efficiently produce multi-scale feature representations in one step (Figure \ref{figoverview}C). We define our multi-scale Fourier convolution as
\begin{equation}
  \left\{ \mathbf{Re}\{\mathcal{F}^{-1}\left\{ \bm{\mathcal{W}_1} \odot \mathrm{crop}_1\left[\bm{\mathfrak{c}}\right] \right\}\}, \ldots, \mathbf{Re}\{\mathcal{F}^{-1}\left\{ \bm{\mathcal{W}_n} \odot \mathrm{crop}_n\left[\bm{\mathfrak{c}}\right] \right\}\} \right\}
\end{equation}
where subscript denotes scale level (higher subscript indicates lower spatial scale/more cropping in Fourier space) and we precompute $\bm{\mathfrak{c}} \coloneqq \mathcal{F}\left \{ \mathbf{c} \right \}$ once.
\setlength{\abovedisplayskip}{7pt plus2pt minus5pt}
\setlength{\belowdisplayskip}{\abovedisplayskip}

\begin{table}[h]
\centering
    \caption{Quality of natural image reconstruction on the DiffuserCam Lensless Mirflickr Dataset (mean \(\pm\) s.e.m., \(n = 999\)). Superscripts denote loss function: \textsuperscript{1} MSE, \textsuperscript{2} MSE+LPIPS.}
\label{tabsupdlmd}
\begin{tabular}{@{}p{27.8mm}@{\hskip 1.0mm}p{23.1mm}@{\hskip 1.0mm}p{17mm}lp{17.5mm}p{17.5mm}@{}}
    \toprule
    \textbf{Method} & \textbf{MSE} \(\downarrow\) (\(\times10^{-2}\)) & \textbf{LPIPS} \(\downarrow\) & \textbf{MS-SSIM} \(\uparrow\) & \textbf{PSNR} \(\uparrow\) & \textbf{Time} \(\downarrow\) (ms)\\
    \midrule
    FourierNet$^1$ & \textbf{0.39} $\pm$ \textbf{0.007} & 0.20 $\pm$ 0.00 & \textbf{0.882} $\pm$ \textbf{0.001} & \textbf{24.8} $\pm$ \textbf{0.09} & 35.54\\
    FourierNet$^2$ & 0.54 $\pm$ 0.010 & \textbf{0.16} $\pm$ \textbf{0.00} & 0.868 $\pm$ 0.001 & 23.4 $\pm$ 0.09 & 35.54\\
    FourierUNet$^1$ & 0.43 $\pm$ 0.009 & 0.22 $\pm$ 0.00 & 0.875 $\pm$ 0.001 & 24.5 $\pm$ 0.09 & 83.63\\
    FourierUNet$^2$ & 0.66 $\pm$ 0.012 & 0.18 $\pm$ 0.00 & 0.853 $\pm$ 0.001 & 22.6 $\pm$ 0.09 & 83.63\\
    \midrule
    Le-ADMM-U$^2$ \cite{Monakhova_Yurtsever_Kuo_Antipa_Yanny_Waller_2019} & 0.75 $\pm$ 0.021 & 0.19 $\pm$ 0.00 & 0.865 $\pm$ 0.002 & 22.1 $\pm$ 0.09 & 48.59\\
    UNet$^2$ \cite{Monakhova_Yurtsever_Kuo_Antipa_Yanny_Waller_2019} & 1.68 $\pm$ 0.060 & 0.24 $\pm$ 0.00 & 0.818 $\pm$ 0.002 & 19.2 $\pm$ 0.11 & \textbf{06.97}\\
    \bottomrule
\end{tabular}
\end{table}

\begin{table}[h]
\centering
\caption{Quality of reconstructed volumes after optimizing microscope parameters to image zebrafish on $256 \times 256$ pixel camera (mean $\pm$ s.e.m., $n = 10$)}
\label{tabsupmetrics}
\begin{tabular}{@{}lllll@{\hskip 1.0mm}c@{}}
    \toprule
    \textbf{Microscope} & \textbf{Reconstruction} & $\bm{L_{\mathrm{HNMSE}}}$ $\downarrow$ & \textbf{MS-SSIM} $\uparrow$ & \textbf{PSNR} $\uparrow$ & \textbf{Time} $\downarrow$ (s)\\
    \midrule
    FourierNet2D & FourierNet3D & \textbf{0.6093} $\pm$ \textbf{0.0209} & \textbf{0.955} $\pm$ \textbf{0.004} & \textbf{34.89} $\pm$ \textbf{0.88} & \textbf{0.38}\\
    FourierNet2D & FourierUNet3D & \textbf{0.5997} $\pm$ \textbf{0.0219} & \textbf{0.956} $\pm$ \textbf{0.003} & \textbf{34.87} $\pm$ \textbf{0.82} & 0.72\\
    Wiener + UNet & Wiener + UNet & 0.7223 $\pm$ 0.0179 & \textbf{0.957} $\pm$ \textbf{0.003} & \textbf{34.49} $\pm$ \textbf{0.91} & 0.73\\
    FourierNet2D & UNet3D & 0.7298 $\pm$ 0.0151 & 0.923 $\pm$ 0.008 & 30.16 $\pm$ 0.94 & 0.96\\
    UNet2D & UNet3D & 0.7109 $\pm$ 0.0161 & 0.913 $\pm$ 0.009 & 29.17 $\pm$ 1.13 & 0.96\\
    \bottomrule
\end{tabular}
\end{table}

\subsection{Global receptive field is more important than multiscale features}
UNets are effective because: (1) features are computed at multiple scales and (2) large receptive fields are achieved in few layers. FourierUNets allowed us to decouple these two explanations because the receptive field is global in a single layer. We see on both our microscopy dataset (which does not have multiscale structure) and the lensless photography dataset (which does have multiscale structure) that the FourierUNet does not improve upon the FourierNet. Thus we see that it is more important for decoding from non-local optical encoders to have a global receptive field than multi-scale features.

\subsection{Training PSFs and volume reconstruction networks}
\label{appendixpsftraining}
\begin{figure}[!h]
  \includegraphics[width=\linewidth]{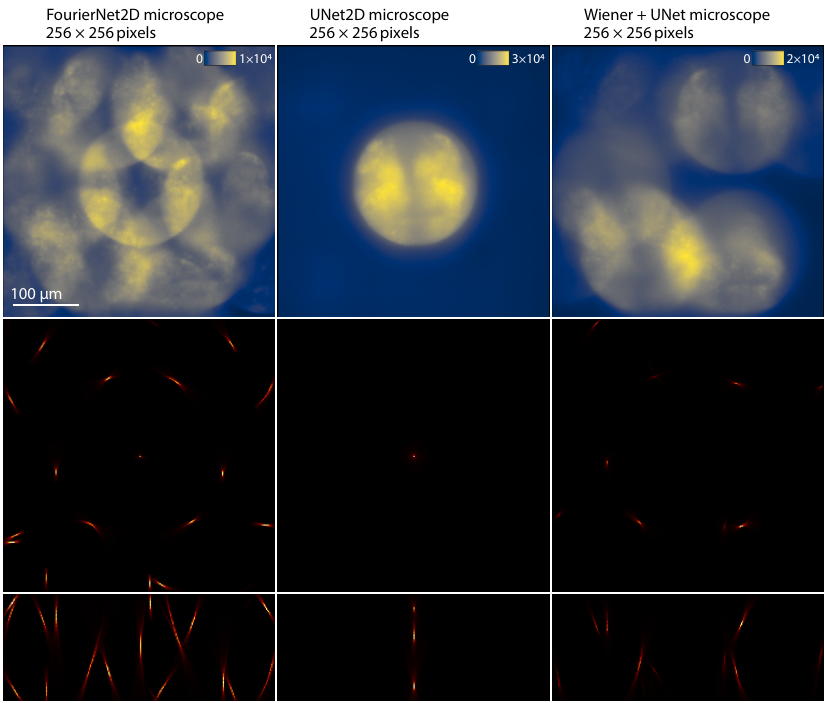}
  \caption{FourierNet successfully optimizes an optical encoder (PSF) to image and reconstruct zebrafish where UNet fails. The FourierNet learned to produce multiple pencils in its optical encoder, which create multiple views of the volume in the camera image. UNet learned only a single pencil and fails to utilize the majority of pixels in the camera image to encode views of the volume. Wiener + UNet produced an optical encoder with multiple pencils, but they do not make as optimal use of the camera pixels as the FourierNet optical encoder. Top row shows simulated camera image of a zebrafish using each optical encoder, middle row shows xy max projection of the optical encoder, and bottom row shows xz max projection of the optical encoder.}
  \label{figsuppsfcomparison}
\end{figure}

\begin{figure}[!h]
  \includegraphics[width=\linewidth]{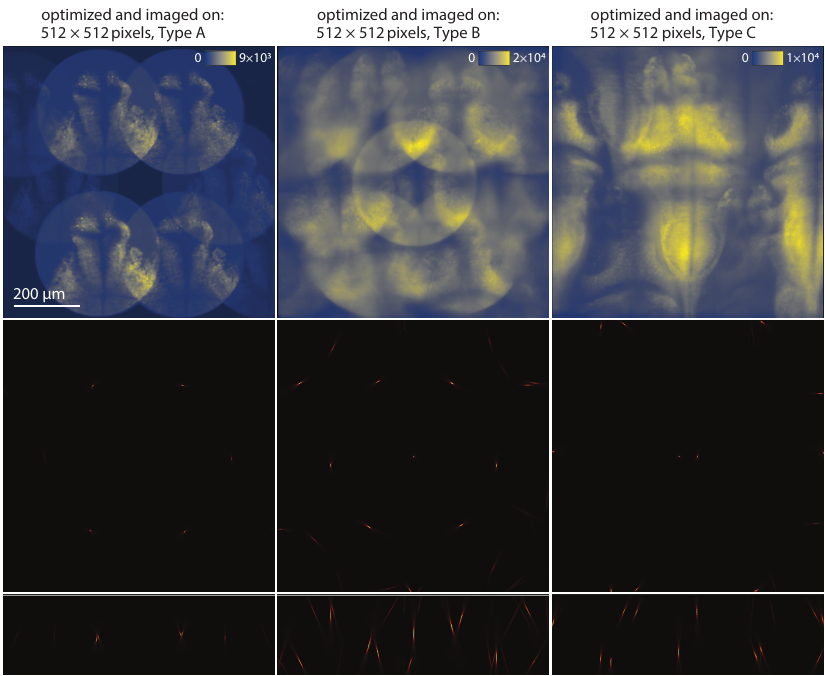}
  \caption{Optimizing optical encoders (PSFs) for different ROIs result in PSFs tailored to each ROI. Note that optical encoder optimized for Type A (left) has pencils with a span in $z$ that matches Type A. Optical encoder optimized for Type B (middle) has pencils that span the entire $z$ depth. Optical encoder optimized for Type C (right) has pencils spread farther apart to account for the larger ROI. Top row shows simulated camera image of a Type A, B, or C example respectively, middle row shows xy max projection of the optical encoder (PSF), and bottom row shows xz max projection of the optical encoder.}
  \label{figsuplargepsfcomparison}
\end{figure}

\begin{figure}[!h]
  \includegraphics[width=\linewidth]{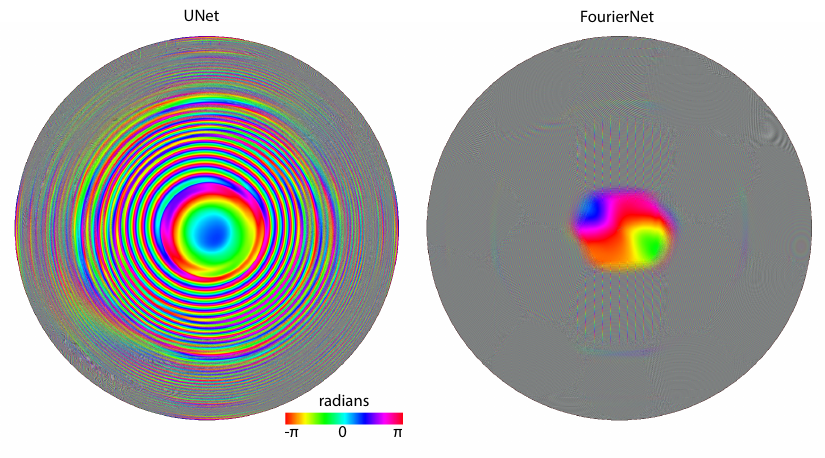}
  \caption{Phase masks (microscope parameters $\bm{\upphi}$) for FourierNet versus UNet. Note that while both phase masks are high-enough frequency to make viewing all pixels difficult after resizing for display and cause their appearance to be gray, the UNet phase mask is much smoother (lower frequency) than the FourierNet phase mask, resulting in a more local optical encoder.}
  \label{figsupunetphase}
\end{figure}

\begin{figure}[!h]
  \includegraphics[width=\linewidth]{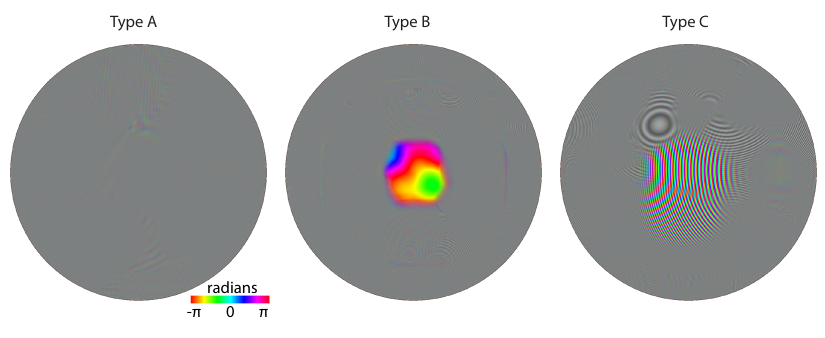}
  \caption{Phase masks (microscope parameters $\bm{\upphi}$) for microscopes optimized for Type A, B, C ROIs, respectively. Note that all phase masks are higher frequency than can be displayed after resizing for this figure, which results in the gray appearance of these phase masks.}
  \label{figsuptypeabcphase}
\end{figure}

Given a simulation of imaging, we can define two modes of autoencoder training: (1) jointly training the phase mask parameters $\bm{\upphi}$ and weak reconstruction networks in order to learn a good optical encoder for a particular class of ROIs (i.e. samples with the same spatiotemporal statistics), and (2) training a stronger reconstruction network only with a fixed, pre-trained $\bm{\upphi}$.

\textbf{Definition of terms}
For both cases of training, the general framework is to simulate imaging using confocal volumes of pan-neuronal labeled larval zebrafish, reconstruct from the simulated image, then update the reconstruction network and, if desired, the microscope parameters. We will define the microscope parameters as $\bm{\upphi}$ and the reconstruction network parameters as $\bm{\uptheta}$ for any reconstruction network $\mathbf{R}_{\bm{\uptheta}}(\mathbf{c})$ where $\mathbf{R}_{\bm{\uptheta}}$ maps 2D images to 3D volume reconstructions. For our training algorithms listed below, we also define: $\mathbf{D}$ our \textbf{dataset}, $\mathbf{v}$ a \textbf{ground truth volume}, $\mathbf{\hat{v}}$ a \textbf{reconstructed volume}, $L$ a computed \textbf{loss}, $z_s$ a list of $z$ plane indices that will be imaged/reconstructed, $\alpha_{\bm{\upphi}}$ the learning rate for the microscope parameters, $\alpha_{\bm{\uptheta}}$ the learning rate for the reconstruction network parameters, and $\beta$ the weight of the non-high pass filtered component of the loss. When selecting a random ground truth volume, we also perform random shift, rotation, flip, and brightness augmentations.

\textbf{Microscope simulation parameters}
When simulating the zebrafish imaging, we use a wavelength of 0.532 $\upmu$m for all simulations. The NA of our microscope is 0.8. The refractive index $n$ is 1.33. We downsample all volumes to (1.0 $\upmu$m z, 1.625 $\upmu$m y, 1.625 $\upmu$m x). We use a taper width of 5 for all simulations, and simulate the optical encoder (PSF) at 50\% larger dimensions in x and y. The resolution of the camera (for all zebrafish datasets) is also (1.625 $\upmu$m y, 1.625 $\upmu$m x).

\textbf{Initialization of} $\bm{\upphi}$
For Type A, B, and our small $256 \times 256$ pixel experiments, we initialize $\bm{\upphi}$ to produce an optical encoder (PSF) consisting of 6 pencil beams at different locations throughout the depth of the volume, with the centers of these beams arranged in a hexagonal pattern in x and y. Because our optimizations generally find optical encoders (PSFs) with many pencils, we find that initializing with such a pattern helps to converge to a more optimal optical encoder (data not shown).

For Type C, we instead initialize with a single helix spanning the depth of the volume (the ``Potato Chip'' from \cite{Broxton_2017}), which seems to find a local minimum for $\bm{\upphi}$ that produces an optical encoder with more pencils (and therefore views in the camera image).

\textbf{Data settings and augmentation for zebrafish data}
Using our total 58 training zebrafish volumes and 10 testing zebrafish volumes (imaged through confocal microscopy), we crop in four different ways to create four different datasets. For training volumes, we crop from random locations from each volume as a form of augmentation. For testing, we crop from the same location. Physically, these crops correspond to either placing a circular aperture before light hits the $4f$ system or changing the illumination thickness in $z$, because samples would be illuminated from the side in a real implementation of this microscope. We model these by cropping cylinders (or cubes if there is no aperture) of different diameters and heights. We show details for all types Type A, B, C in Table \ref{tabdatasets}, where the diameter of the cylinder is labeled ``Aperture Diameter'' and the illumination thickness is labeled ``Height''. For our small initial experiments to compare UNets and FourierNets for optimizing phase masks, we simulated a camera with $256 \times 256$ pixels and during reconstruction each volume had 96 planes, a field of view of (200 $\upmu$m z, 416 $\upmu$m y, 416 $\upmu$m x), and a cylindrical cutout diameter of 193 $\upmu$m.

We augment our volumes during training by taking random locations from these volumes, randomly flipping the volumes in both z and y, and also randomly rotating in pitch, yaw, and roll. Most importantly, we also randomly scale the brightness of our samples and add random background levels which serve to adjust the signal-to-noise ratio (SNR) of the resulting simulated images. The only exception to these augmentations is Type C, where we set all the volumes to the same in-plane vertical orientation (while still applying rotation augmentations in pitch and roll).

\begin{table}[!h]
    \centering
    \caption{Specifications of all zebrafish datasets Type A, B, C for reconstruction}
    \label{tabdatasets}
    \begin{tabular}{@{}ccccc@{}}
         \toprule
         \textbf{Dataset} & \textbf{Camera} (px) &  \textbf{Height (planes)} & \textbf{Span (z, y, x)} ($\upmu$m) & \textbf{Aperture Diameter} ($\upmu$m)\\
         \midrule
         Type A & $512 \times 512$ & 12 & (25, 832, 832) & 386\\\midrule
         Type B & $512 \times 512$ & 128 & (250, 832, 832) & 386\\\midrule
         Type C & $512 \times 512$ & 128 & (250, 832, 832) & -\\\midrule
         \bottomrule
    \end{tabular}
\end{table}

\textbf{Parallelizing imaging and reconstruction}
\begin{figure}[!h]
  \includegraphics[width=\linewidth]{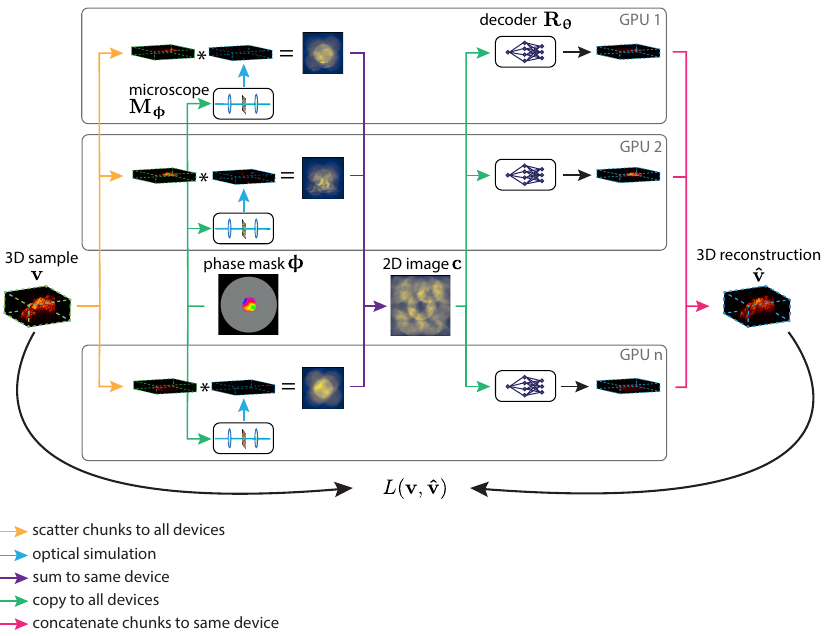}
  \caption{Overview of our parallelization strategy, visualizing Algorithm \ref{algparallelpsftraining}. Colored arrows describe parallelization communication, reduction, or computation. The volume is split into chunks, which are scattered to all compute devices (GPUs). Gray boxes demarcate each independent chunk being processed in parallel on its own device. Optical encoder (PSF) computation is simulated for each chunk in parallel, followed by parallel simulation of the imaging as a convolution of the optical encoder and sample chunk. Then, the partial images are summed onto the same device (GPU 1). After simulating shot noise on this single device, the final image is copied to all devices and each chunk is reconstructed on each device in parallel. At this point the loss can be partially computed in parallel for each chunk and then summed for the loss between the full volumes, or the chunked reconstruction can be concatenated to the same device. We show the loss computation at once rather than in parallel for simplification of visualization only. Contrasts for volumes and optical encoders in chunks have been artificially boosted for visibility. Zoom in for best viewing.}
  \label{figsupparallelsim}
\end{figure}

We show our parallelization strategy for both imaging and reconstruction in Figure \ref{figsupparallelsim} as well as in the following algorithms. Because this simulation can become too expensive in memory to fit on a single device, we generally perform the simulation, reconstruction, and loss calculation in parallel for both training modes. Therefore, any variable that has a $_s$ subscript refers to a list of chunks of that variable that will be run on each device. A $^j$ superscript indicates a particular chunk for GPU $j$. For example $z_s$ is a list of plane indices to be imaged/reconstructed, and $z_s^j$ is the $j^{\mathrm{th}}$ chunk of plane indices that will be imaged/reconstructed on GPU $j$. We denote \texttt{parallel} for any operations that are performed in parallel and \texttt{scatter} for splitting data into chunks and spreading across multiple GPUs. Imaging can be cleanly parallelized: chunks of an optical encoder (PSF) and sample can be partially imaged on multiple GPUs independently because the convolution occurs per plane, then finally all partial images can be summed together onto a single GPU. The reconstructions can similarly take the final image and reconstruct partial chunks (as well as calculate losses on partial chunks) of the volume independently per device. We implicitly gather data to the same GPU when computing sums ($\sum$) or means ($\mathbb{E}$). The functions \texttt{parallel image} and \texttt{compute PSF} follow the definitions above in equations \ref{eqimaging} and \ref{eqpsf}. In the algorithms shown here, \texttt{parallel image} applies the same convolution described above in equation \ref{eqimaging}.

\textbf{Sparse gradients and downsampling}
We additionally support training and reconstructing only some of the planes for imaging and potentially a different partial group of planes during reconstruction, as a way to sparsely compute gradients for optimization of $\theta_m$ and save memory. The planes not imaged with gradients can still contribute to the image (without their gradients being tracked) in order to make the problem more difficult for the reconstruction network. Over multiple iterations, this can become equivalent to the more expensive densely computed gradient method, essentially trading training time for memory. An additional memory saving measure not written in the algorithms is to compute the optical encoder (PSF) at a high resolution, then downsample the optical encoder using a 2D sum pool to preserve total energy in order to reduce memory usage when performing the imaging and reconstruction. We denote \textbf{with no gradient tracking} to show an operation without gradients.

\subsection{Comparing number of optical encoder voxels in simulation to previous works}
\label{appendixvoxelcomparison}
We compare our optical encoders (PSFs), which are simulated at a maximum size of $64 \times 3840 \times 3840$ voxels in z, y, and x (prior to downsampling for the simulation of imaging) respectively to those of deep learning optical encoder optimizations in localization microscopy and depth from defocus \cite{Nehme_Freedman_Gordon_Ferdman_Weiss_Alalouf_Naor_Orange_Michaeli_Shechtman_2020, Ikoma_Nguyen_Metzler_Peng_Wetzstein_2021}. We simulate at such high voxel counts in order to allow the phase mask (at a size of $3840 \times 3840$ voxels) to produce high frequencies, which are required for producing features in the optical encoder near the edges of the field of view. For localization microscopy, the optical encoders are $51 \times 329 \times 329$ voxels, which means our optical encoder has approximately 170$\times$ more voxels \cite{Nehme_Freedman_Gordon_Ferdman_Weiss_Alalouf_Naor_Orange_Michaeli_Shechtman_2020}. For a state of the art depth from defocus implementation, the optical encoder is simulated with rotational symmetry, which means the actual simulation can occur in only one dimension per depth plane \cite{Ikoma_Nguyen_Metzler_Peng_Wetzstein_2021}. Thus, the depth from defocus optical encoder is simulated with $16 \times 8000$ voxels in z and x, respectively \cite{Ikoma_Nguyen_Metzler_Peng_Wetzstein_2021}. The single dimension along x is then rotated to produce the full optical encoder at each of the 16 depth planes \cite{Ikoma_Nguyen_Metzler_Peng_Wetzstein_2021}. This means our optical encoder has approximately 7372$\times$ more unique voxels.

\begin{algorithm}[!h]
  \DontPrintSemicolon
  \SetKwInOut{Input}{Input}
  \SetKw{KwFor}{for}
  \SetKw{KwIn}{in}
  \SetKw{KwUsing}{using}
  \SetKwFunction{Shuffle}{shuffle}
  \SetKwFunction{Scatter}{scatter}
  \SetKwFunction{Sum}{sum}
  \SetKwFunction{Mean}{mean}
  \SetKwFunction{HighPass}{high pass}
  \SetKwFunction{Parallel}{parallel}
  \SetKwFunction{Image}{image}
  \SetKwFunction{ComputePSF}{compute PSF}
  \SetKwFunction{Update}{update}
  \SetKwFunction{Grad}{compute gradient}
  \SetKwFunction{ComputeLoss}{compute loss}
  \SetKwFunction{SelectPlanes}{select planes}
  \SetKwFunction{Adam}{\textsc{Adam}}
  \SetKwBlock{NoGrad}{with no gradient tracking}{end}
  \SetKwFunction{ParIm}{parallel image}
  \SetKwFunction{ParRecLoss}{parallel reconstruct/loss}
  \Input{$\mathbf{M}_{\bm{\upphi}}, \bm{\upphi}, \alpha_{\bm{\upphi}}, \mathbf{R}_{\bm{\uptheta}}, \bm{\uptheta}, \alpha_{\bm{\uptheta}}, \mathbf{D}, z_s, \beta$}
  \For{$\mathbf{v} \in \mathbf{D}$}{
    \tcp{select plane indices to be imaged with and without gradients}
    $z_{s, \mathrm{gradient}}, z_{s, \mathrm{no~gradient}} \leftarrow$ \SelectPlanes{$z_s$}\;
    \tcp{move sample planes to be imaged with gradients to multiple GPUs}
    $\mathbf{v}_{s, \mathrm{gradient}} \leftarrow$ \Scatter{$\mathbf{v}, z_{s, \mathrm{gradient}}$}\;
    \tcp{move sample planes to be imaged without gradients to multiple GPUs}
    $\mathbf{v}_{s, \mathrm{no~gradient}} \leftarrow$ \Scatter{$\mathbf{v}, z_{s, \mathrm{no~gradient}}$}\;
    \tcp{compute PSF with gradients on multiple GPUs}
    $\mathbf{s}_{s, \mathrm{gradient}} \leftarrow$ \Parallel{\ComputePSF{$\mathbf{M}_{\bm{\upphi}}, z_s^j$} \KwFor $z_s^j$ \KwIn $z_{s, \mathrm{gradient}}$}\;
    \tcp{compute partial image with gradients on multiple GPUs}
    $\mathbf{c}_{\mathrm{gradient}} \leftarrow$ \ParIm{$\mathbf{s}_{s, \mathrm{gradient}}, \mathbf{v}_{s, \mathrm{gradient}}$}\;
    \tcp{compute PSF without gradients on multiple GPUs}
    \NoGrad{$\mathbf{s}_{s, \mathrm{no~gradient}} \leftarrow$ \Parallel{\ComputePSF{$\mathbf{M}_{\bm{\upphi}}, z_s^j$} \KwFor $z_s^j$ \KwIn $z_{s, \mathrm{no~gradient}}$}}
    \tcp{compute partial image without gradients on multiple GPUs}
    \NoGrad{$\mathbf{c}_{\mathrm{no~gradient}} \leftarrow$ \ParIm{$\mathbf{s}_{s, \mathrm{no~gradient}}, \mathbf{v}_{s, \mathrm{no~gradient}}$}}
    \tcp{compute full image by summing partial images onto one GPU}
    $\mathbf{c} \leftarrow$ $\sum[\mathbf{c}_{\mathrm{gradient}}, \mathbf{c}_{\mathrm{no~gradient}}]$\;
    \tcp{select plane indices to be reconstructed}
    $z_{s, \mathrm{reconstruct}} \leftarrow$ \SelectPlanes{$z_s$}\;
    \tcp{move sample planes that will be reconstructed to multiple GPUs}
    $\mathbf{v}_{s, \mathrm{reconstruct}} \leftarrow$ \Scatter{$\mathbf{v}, z_{s, \mathrm{reconstruct}}$}\;
    \tcp{compute mean of high passed sample for loss normalization}
    $\mu_{H(\mathbf{v})} \leftarrow \mathbb{E}(H(\mathbf{v}_{s, \mathrm{reconstruct}})^2)$\;
    \tcp{compute mean of sample for loss normalization}
    $\mu_{\mathbf{v}} \leftarrow \mathbb{E}(\mathbf{v}_{s, \mathrm{reconstruct}}^2)$\;
    \tcp{move reconstruction networks to multiple GPUs}
    $\mathbf{R}_{\bm{\uptheta}, s} \leftarrow$ \Scatter{$\mathbf{R}_{\bm{\uptheta}}$} \;
    \tcp{compute reconstruction and loss on multiple GPUs}
    $L \leftarrow$ \ParRecLoss{$\mathbf{c}, \mathbf{v}_{s, \mathrm{reconstruct}}, \mathbf{R}_{\bm{\uptheta}, s}, \mu_{H(\mathbf{v})}, \mu_{\mathbf{v}}, \beta$}\;
    \tcp{compute gradients for all parameters}
    $g_{\bm{\uptheta}} \leftarrow \nabla_{\bm{\uptheta}} L$\;
    $g_{\bm{\upphi}} \leftarrow \nabla_{\bm{\upphi}} L$\;
    \tcp{update all parameters}
    $\bm{\uptheta} \leftarrow $\Adam{$\alpha_{\bm{\uptheta}},\bm{\uptheta},g_{\bm{\uptheta}}$}\;
    $\bm{\upphi} \leftarrow $\Adam{$\alpha_{\bm{\upphi}},\bm{\upphi},g_{\bm{\upphi}}$}\;
    }
  \caption{Parallel optical encoder (PSF) engineering by joint training of reconstruction network and phase mask. Microscope $\mathbf{M}_{\bm{\upphi}}$ parameters are $\bm{\upphi}$, reconstruction network $\mathbf{R}_{\bm{\uptheta}}$ parameters are $\bm{\uptheta}$, dataset is $\mathbf{D}$, learning rates for $\bm{\upphi}$ and $\bm{\uptheta}$ are $\alpha_{\bm{\upphi}}$ and $\alpha_{\bm{\uptheta}}$ respectively, plane indices to image and reconstruct from $z_s$, and weight for $L_{\mathrm{NMSE}}$ is $\beta$.}
  \label{algparallelpsftraining}
\end{algorithm}

\begin{algorithm}[!h]
  \DontPrintSemicolon
  \SetKwInOut{Input}{Input}
  \SetKw{KwFor}{for}
  \SetKw{KwIn}{in}
  \SetKw{KwUsing}{using}
  \SetKwFunction{Shuffle}{shuffle}
  \SetKwFunction{Scatter}{scatter}
  \SetKwFunction{Sum}{sum}
  \SetKwFunction{Mean}{mean}
  \SetKwFunction{HighPass}{high pass}
  \SetKwFunction{Parallel}{parallel}
  \SetKwFunction{Image}{image}
  \SetKwFunction{ComputePSF}{compute PSF}
  \SetKwFunction{Update}{update}
  \SetKwFunction{Grad}{compute gradient}
  \SetKwFunction{ComputeLoss}{compute loss}
  \SetKwFunction{Concatenate}{concatenate}
  \SetKwFunction{SelectPlanes}{select planes}
  \SetKwFunction{Adam}{\textsc{Adam}}
  \SetKwBlock{NoGrad}{with no gradient tracking}{end}
  \SetKwFunction{ParIm}{parallel image}
  \SetKwFunction{ParRecLoss}{parallel reconstruct/loss}
  \Input{$\mathbf{M}_{\bm{\upphi}}, \bm{\upphi}, \alpha_{\bm{\upphi}}, \mathbf{R}_{\bm{\uptheta}}, \bm{\uptheta}, \alpha_{\bm{\uptheta}}, \mathbf{D}, z_s, \beta$}
  \tcp{compute PSF without gradients on multiple GPUs}
  \NoGrad{$\mathbf{s}_{\mathrm{no~gradient}} \leftarrow ${\Parallel{\ComputePSF{$\mathbf{M}_{\bm{\upphi}}, z_s^j$} \KwFor $z_s^j$ \KwIn $z_{s}$}}}
  \For{$\mathbf{v} \in \mathbf{D}$}{
    \tcp{select plane indices to be imaged without gradients}
    $z_{s, \mathrm{no~gradient}} \leftarrow$ \SelectPlanes{$z_s$}\;
    \tcp{move sample planes to be imaged without gradients to multiple GPUs}
    $\mathbf{v}_{s, \mathrm{no~gradient}} \leftarrow$ \Scatter{$\mathbf{v}, z_{s, \mathrm{no~gradient}}$}\;
    \tcp{move necessary PSF planes to multiple GPUs}
    $\mathbf{s}_{s, \mathrm{no~gradient}} \leftarrow$ \Scatter{$\mathbf{s}_{\mathrm{no~gradient}}, z_{s, \mathrm{no~gradient}}$}\;
    \tcp{compute image without gradients on multiple GPUs}
    \NoGrad{$\mathbf{c} \leftarrow$ \ParIm{$\mathbf{s}_{s, \mathrm{no~gradient}}, \mathbf{v}_{s, \mathrm{no~gradient}}$}}
    \tcp{select plane indices to be reconstructed}
    $z_{s, \mathrm{reconstruct}} \leftarrow$ \SelectPlanes{$z_s$}\;
    \tcp{move sample planes that will be reconstructed to multiple GPUs}
    $\mathbf{v}_{s, \mathrm{reconstruct}} \leftarrow$ \Scatter{$\mathbf{v}, z_{s, \mathrm{reconstruct}}$}\;
    \tcp{compute mean of high passed sample for loss normalization}
    $\mu_{H(\mathbf{v})} \leftarrow \mathbb{E}[H(\mathbf{v}_{s, \mathrm{reconstruct}})^2]$\;
    \tcp{compute mean of sample for loss normalization}
    $\mu_{\mathbf{v}} \leftarrow \mathbb{E}[\mathbf{v}_{s, \mathrm{reconstruct}}^2]$\;
    \tcp{move reconstruction networks to multiple GPUs}
    $\mathbf{R}_{\bm{\uptheta}, s} \leftarrow$ \Scatter{$\mathbf{R}_{\bm{\uptheta}}$}\;
     \tcp{compute reconstruction and loss on multiple GPUs}
    $L \leftarrow$ \ParRecLoss{$\mathbf{c}, \mathbf{v}_{s, \mathrm{reconstruct}}, \mathbf{R}_{\bm{\uptheta}, s}, \mu_{H(\mathbf{v})}, \mu_{\mathbf{v}}, \beta$}\;
    \tcp{compute gradients for reconstruction networks only}
    $g_{\bm{\uptheta}} \leftarrow \nabla_{\bm{\uptheta}} L$\;
    \tcp{update reconstruction network parameters only}
    $\bm{\uptheta} \leftarrow $\Adam{$\alpha_{\bm{\uptheta}},\bm{\uptheta},g_{\bm{\uptheta}}$}\;
    }
  \caption{Parallel training a reconstruction network given a pre-trained phase mask. Microscope $\mathbf{M}_{\bm{\upphi}}$ parameters are $\bm{\upphi}$ (phase mask), reconstruction network $\mathbf{R}_{\bm{\uptheta}}$ parameters are $\bm{\uptheta}$, dataset is $\mathbf{D}$, learning rates for $\bm{\upphi}$ and $\bm{\uptheta}$ are $\alpha_{\bm{\upphi}}$ and $\alpha_{\bm{\uptheta}}$ respectively, plane indices to image and reconstruct from are $z_s$, and weight for $L_{\mathrm{NMSE}}$ is $\beta$.}
\end{algorithm}

\begin{algorithm}[!h]
  \DontPrintSemicolon
  \SetKwInOut{Input}{Input}
  \SetKwInOut{Output}{Output}
  \SetKw{KwFor}{for}
  \SetKw{KwIn}{in}
  \SetKw{KwUsing}{using}
  \SetKwFunction{Shuffle}{shuffle}
  \SetKwFunction{Scatter}{scatter}
  \SetKwFunction{Sum}{sum}
  \SetKwFunction{Mean}{mean}
  \SetKwFunction{HighPass}{high pass}
  \SetKwFunction{Parallel}{parallel}
  \SetKwFunction{Convolve}{convolve}
  \SetKwFunction{ComputePSF}{compute PSF}
  \SetKwFunction{Update}{update}
  \Input{$\mathbf{s}_s, \mathbf{v}_s$}
  \Output{$\mathbf{c}$}
  \tcp{compute images in parallel on multiple GPUs, then sum to single GPU}
  $\mathbf{c} \leftarrow$ $\sum[$\Parallel{\Convolve{$\mathbf{s_s^j}, \mathbf{v_s^j}$} \KwFor $(\mathbf{s}_s^j, \mathbf{v_s^j})$ \KwIn $(\mathbf{s}_s, \mathbf{v}_s)$}$]$\;
  \Return{$\mathbf{c}$}\;
  \caption{Parallel imaging. Optical encoder (PSF) planes on multiple GPUs are $\mathbf{s}_s$, sample planes on multiple GPUs to be imaged are $\mathbf{v}_s$.}
\end{algorithm}

\begin{algorithm}[!h]
  \DontPrintSemicolon
  \SetKwInOut{Input}{Input}
  \SetKwInOut{Output}{Output}
  \SetKw{KwFor}{for}
  \SetKw{KwIn}{in}
  \SetKw{KwUsing}{using}
  \SetKwFunction{Shuffle}{shuffle}
  \SetKwFunction{Scatter}{scatter}
  \SetKwFunction{Sum}{sum}
  \SetKwFunction{Mean}{mean}
  \SetKwFunction{HighPass}{high pass}
  \SetKwFunction{Parallel}{parallel}
  \SetKwFunction{Image}{image}
  \SetKwFunction{ComputePSF}{compute PSF}
  \SetKwFunction{Concatenate}{concatenate}
  \SetKwFunction{Update}{update}
  \Input{$\mathbf{c}, \mathbf{v}_s, \mathbf{R}_{\bm{\uptheta}, s}, \mu_{H(\mathbf{v})}, \mu_{\mathbf{v}}, \beta$}
  \Output{L}
  \tcp{compute reconstruction and loss in parallel on multiple GPUs}
  $\mathbf{\hat{v}}_s \leftarrow$ \Concatenate{\Parallel{$\mathbf{R}_s^j(\mathbf{c})$ \KwFor $\mathbf{R}_s^j$ \KwIn $\mathbf{R}_{\bm{\uptheta}, s}$}}\;
  $L_s \leftarrow$ \Parallel{$\frac{\mathbb{E}[(H(\mathbf{v}_s^j) - H(\mathbf{\hat{v}}_s^j))^2]}{\mu_{H(\mathbf{v})}} + \beta\frac{\mathbb{E}[(\mathbf{v}_s^j - \mathbf{\hat{v}}_s^j)^2]}{\mu_{\mathbf{v}}}$ \KwFor $(\mathbf{\hat{v}}_s^j, \mathbf{v}_s^j)$ \KwIn $(\mathbf{\hat{v}}_s, \mathbf{v}_s)$}\;
  \tcp{compute mean of scattered losses on single GPU}
  $L \leftarrow \mathbb{E}[L_s]$\;
  \Return{$L$}\;
  \caption{Parallel reconstruction/loss calculation. Camera image is $\mathbf{c}$, sample planes on multiple GPUs are $\mathbf{v}_s$, reconstruction networks on multiple GPUs are $\mathbf{R}_{\bm{\uptheta}, s}$, mean for $L_{\mathrm{HNMSE}}$ normalization is $\mu_{H(\mathbf{v})}$, mean for $L_{\mathrm{NMSE}}$ normalization is $\mu_{\mathbf{v}}$, and weight for $L_{\mathrm{NMSE}}$ is $\beta$.}
\end{algorithm}

\subsection{Implementation details}
\label{appendiximplementationdetails}
\textbf{Fourier convolution details}
Our Fourier convolution uses complex number weights, implemented as two channels of real numbers. Furthermore, in order to prevent the convolution from wrapping around the edges, we have to pad the input to double the size. The size of the weight must match the size of this padded input. This means that the number of parameters for our Fourier convolution implementation is $8\times$ the number of parameters required for a global kernel in a spatial convolution (though the Fourier convolution is significantly faster). We do this to save an extra padding and Fourier operation, trading memory for speed. Because the simulation of imaging requires more memory than the reconstruction network, we found this to be an acceptable tradeoff.

\textbf{Common network details}
All convolutions (including Fourier convolutions) use ``same'' padding. For FourierUNets and vanilla UNets, downsampling and upsampling is performed only in the $x$ and $y$ dimensions (we do not downsample or upsample in $z$ because there could potentially not be enough planes to do so). We train all networks using the \textsc{adam} optimizer with all default PyTorch parameters except the learning rate, which we always set to $10^{-4}$ for the reconstruction network parameters $\bm{\uptheta}$ and $10^{-2}$ for the phase mask parameters $\bm{\upphi}$.

\textbf{Normalization}
We use \textbf{input scaling} during both training and inference in order to normalize out differences in the brightness of the image and prevent instabilities in our gradients. This means we divide out the median value of the input (scaled by some factor in order to bring the loss to a reasonable range) and then undo this scaling after the output of the network. This effectively linearizes our reconstruction networks, meaning a scaling of the image sent to the network will exactly scale the output by that value. We also find this is a more effective and simpler alternative to using a BatchNorm on our inputs. We continue to use BatchNorm between our convolution layers within the reconstruction network \cite{Ioffe_Szegedy_2015}, which is effectively InstanceNorm in our case where batch size is 1 \cite{Ulyanov_Vedaldi_Lempitsky_2017}.

\textbf{Planewise network training logic}
When we train optical encoders by optimizing $\bm{\upphi}$, we train separate reconstruction networks per plane. This allows us to flexibly compute sparse gradients across different planes from iteration to iteration, as described in Appendix \ref{appendixpsftraining}. In order to do this, we create placeholder networks on any number of GPUs, then copy the parameters stored on CPU for each plane's reconstruction network to a network on the GPU as needed during a forward pass. After calculating an update with the optimizer, we copy the parameter values back to the corresponding parameter on CPU.

\textbf{Training times}
We optimize our smaller, $256 \times 256$ pixel microscopy experiments on 4 RTX 2080 Ti GPUs when optimizing both $\bm{\upphi}$ and $\bm{\uptheta}$ and 8 RTX 2080 Ti GPUs when optimizing only $\bm{\uptheta}$, except the Wiener + UNet model which is trained on 8 RTX Quadro 8000 GPUs. For these, we can compare training times for the different network architectures. One training iteration (including microscope simulation, reconstruction, backpropagation, and parameter update) takes $\sim$0.6 seconds for FourierNet2D, $\sim$1.3 seconds for UNet2D, and $\sim$0.8 seconds for Wiener + UNet when optimizing both $\bm{\upphi}$ and $\bm{\uptheta}$. One training iteration takes $\sim$0.4 seconds for FourierNet3D, $\sim$0.7 seconds for FourierUNet3D, and $\sim$0.8 seconds for UNet3D when only optimizing $\bm{\uptheta}$. Our larger Type A, B, C experiments are always optimized on 8 RTX Quadro 8000 GPUs. More details are found in Tables \ref{tabzdtrainingdetails} and \ref{tabzabctrainingdetails}.

\textbf{Memory usage}
We show our training GPU memory usage for all kinds of snapshot microscopy experiments training both optical encoders and training reconstruction networks only in Table \ref{tabmemoryusage}. Because we must synchronize some computations to a single GPU, there will be one GPU with higher memory usage than the rest. Thus, we report both the highest memory usage of a single GPU (the maximum memory usage across GPUs) as well as the memory usage of the remaining single GPUs (the mode memory usage across GPUs).

\begin{table}[!h]
\centering
\caption{Small experiment training times}
\label{tabzdtrainingdetails}
\begin{tabular}{@{}llrccc@{}}
    \toprule
    \textbf{Network} & \textbf{Optimizing} & \textbf{\# parameters} & \textbf{\# train steps} & \textbf{Train step time} (s) & \textbf{Total time} (h)\\
    \midrule
    FourierNet2D & $\bm{\uptheta}, \bm{\upphi}$ & $\sim4.2 \times 10^7$ & $10^6$ & $\sim 0.8$ & $\sim 222$\\
    FourierNet3D & $\bm{\uptheta}$ & $\sim6.3 \times 10^7$ & $10^6$ & $\sim 0.4$ & $\sim 111$\\
    FourierUNet3D & $\bm{\uptheta}$ & $\sim8.4 \times 10^7$ & $10^6$ & $\sim 0.7$ & $\sim 194$\\
    \midrule
    UNet2D & $\bm{\uptheta}, \bm{\upphi}$ & $\sim4.0 \times 10^7$ & $10^6$ & $\sim 1.3$ & $\sim 361$\\
    Wiener + UNet & $\bm{\uptheta}, \bm{\upphi}$ & $\sim8.0 \times 10^7$ & $5 \times 10^5$ & $\sim 0.8$ & $\sim 111$\\
    UNet3D & $\bm{\uptheta}$ & $\sim1.0 \times 10^8$ & $10^6$ & $\sim 0.8$ & $\sim 222$\\
    \bottomrule
\end{tabular}
\end{table}

\begin{table}[!h]
\centering
\caption{Type A, B, C experiment training times}
\label{tabzabctrainingdetails}
\begin{tabular}{@{}l@{\hskip 0.08in}l@{\hskip 0.01in}r@{\hskip 0.05in}c@{\hskip 0.05in}r@{\hskip 0.05in}c@{\hskip 0.05in}l@{}}
    \toprule
    \textbf{Network} & \textbf{Optimizing} & \textbf{\# parameters} & \textbf{Type} & \textbf{\# train steps} & \textbf{Train step time} (s) & \textbf{Total time} (h)\\
    \midrule
    FourierNet2D & $\bm{\uptheta}, \bm{\upphi}$ & $\sim1.7 \times 10^8$ & A & $5.8 \times 10^5$ & $\sim 1.1$ & $\sim 177$\\
    FourierNet3D & $\bm{\uptheta}$ (fixed $\bm{\upphi}$ for A) & $\sim3.4 \times 10^8$ & A & $\sim 2.6 \times 10^5$ & $\sim 1.6$ & $\sim 116$\\
    FourierNet3D & $\bm{\uptheta}$ (fixed $\bm{\upphi}$ for A) & $\sim3.4 \times 10^8$ & B & $\sim 1.3 \times 10^5$ & $\sim 1.6$ & $\sim 58$\\
    FourierNet3D & $\bm{\uptheta}$ (fixed $\bm{\upphi}$ for A) & $\sim3.4 \times 10^8$ & C & $\sim 1.3 \times 10^5$ & $\sim 1.6$ & $\sim 58$\\
    \midrule
    FourierNet2D & $\bm{\uptheta}, \bm{\upphi}$ & $\sim1.7 \times 10^8$ & B & $5.8 \times 10^5$ & $\sim 1.1$ & $\sim 177$\\
    FourierNet3D & $\bm{\uptheta}$ (fixed $\bm{\upphi}$ for B) & $\sim3.4 \times 10^8$ & A & $\sim 1.2 \times 10^5$ & $\sim 1.6$ & $\sim 53$\\
    FourierNet3D & $\bm{\uptheta}$ (fixed $\bm{\upphi}$ for B) & $\sim3.4 \times 10^8$ & B & $10^6$ & $\sim 1.6$ & $\sim 444$\\
    FourierNet3D & $\bm{\uptheta}$ (fixed $\bm{\upphi}$ for B) & $\sim3.4 \times 10^8$ & C & $\sim 5.0 \times 10^5$ & $\sim 1.6$ & $\sim 222$\\
    \midrule
    FourierNet2D & $\bm{\uptheta}, \bm{\upphi}$ & $\sim1.7 \times 10^8$ & C & $5.8 \times 10^5$ & $\sim 1.1$ & $\sim 177$\\
    FourierNet3D & $\bm{\uptheta}$ (fixed $\bm{\upphi}$ for C) & $\sim3.4 \times 10^8$ & A & $\sim 3.4 \times 10^5$ & $\sim 1.6$ & $\sim 151$\\
    FourierNet3D & $\bm{\uptheta}$ (fixed $\bm{\upphi}$ for C) & $\sim3.4 \times 10^8$ & B & $\sim 3.4 \times 10^5$ & $\sim 1.6$ & $\sim 151$\\
    FourierNet3D & $\bm{\uptheta}$ (fixed $\bm{\upphi}$ for C) & $\sim3.4 \times 10^8$ & C & $\sim 3.7 \times 10^5$ & $\sim 1.6$ & $\sim 164$\\
    \bottomrule
\end{tabular}
\end{table}

\begin{table}[!h]
\centering
\caption{GPU memory usage for all snapshot microscopy experiment types}
\label{tabmemoryusage}
\begin{tabular}{@{}llllrrr@{}}
    \toprule
    \textbf{Network} & \textbf{Optimizing} & \textbf{Type} & \textbf{\# GPUs} & \textbf{Max} (MB) & \textbf{Mode} (MB) & \textbf{Total} (MB)\\
    \midrule
    FourierNet2D & $\bm{\uptheta}, \bm{\upphi}$ & Small & 4 & 4,815 & 4,783 & 19,164\\
    UNet2D & $\bm{\uptheta}, \bm{\upphi}$ & Small & 4 & 5,177 & 5,145 & 20,612\\
    Wiener + UNet & $\bm{\uptheta}, \bm{\upphi}$ & Small & 4 & 6,465 & 6,253 & 50,860\\
    \midrule
    FourierNet3D & $\bm{\uptheta}$ & Small & 8 & 2,647 & 1,617 & 13,966\\
    FourierUNet3D & $\bm{\uptheta}$ & Small & 8 & 2,725 & 1,631 & 14,142\\
    UNet3D & $\bm{\uptheta}$ & Small & 8 & 2,751 & 1,603 & 13,972\\
    \midrule
    FourierNet2D & $\bm{\uptheta}, \bm{\upphi}$ & A, B, C & 8 & 8,513 & 8,267 & 66,382\\
    \midrule
    FourierNet3D & $\bm{\uptheta}$ & A, B, C & 8 & 15,537 & 3,865 & 42,592\\
    \bottomrule
\end{tabular}
\end{table}

\subsection{Details for FourierNets outperform state-of-the-art for reconstructing natural images captured by DiffuserCam lensless camera}
\label{appendixdlmd}
We performed no augmentations for this set of trainings reconstructing RGB color images of natural scenes from RGB diffused images taken through a DiffuserCam \cite{Monakhova_Yurtsever_Kuo_Antipa_Yanny_Waller_2019}. We modified our FourierNet2D architecture to create the FourierNetRGB architecture and our FourierUNet2D architecture to create the FourierUNetRGB architecture, outlined in Table \ref{tabfouriernetrgb} and Table \ref{tabfourierunetrgb} respectively. Training details are shown in Table \ref{tabdlmdtrainingdetails}. Because these reconstructions are of 2D images only and required no microscope simulation, we were able to use a batch size of 4 images per iteration.

\begin{figure}[!h]
  \includegraphics[width=\linewidth]{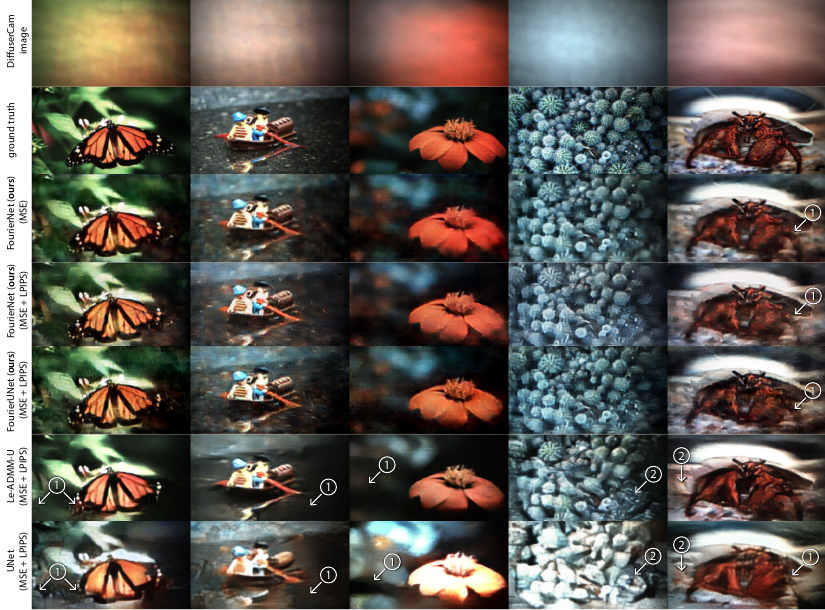}
  \caption{Comparisons of our method (second and third rows) to state-of-the-art learned reconstruction methods on lensless diffused images of natural scenes. Note that FourierNet trained with MSE only shows comparable visual results to training with MSE + LPIPS. Regions labeled {\Large \textcircled{\normalsize 1}} indicate missing details, either resolution or textures in backgrounds. Regions labeled {\Large \textcircled{\normalsize 2}} indicate hallucinated textures. Note that the previous state-of-the-art solutions \cite{Monakhova_Yurtsever_Kuo_Antipa_Yanny_Waller_2019} exhibit both issues more often compared to our models.}
  \label{figsupdlmdcomparison}
  \vspace{-0.2cm}
\end{figure}

\begin{table}[!h]
\centering
\caption{DLMD experiment training times. Superscripts denote loss function: \textsuperscript{1} MSE, \textsuperscript{2} MSE+LPIPS.}
\label{tabdlmdtrainingdetails}
\begin{tabular}{@{}llrccc@{}}
    \toprule
    \textbf{Network} & \textbf{Optimizing} & \textbf{\# parameters} & \textbf{\# train steps} & \textbf{Train step time} (s) & \textbf{Total time} (h)\\
    \midrule
    FourierNetRGB$^1$ & $\bm{\uptheta}$ & $\sim1.6 \times 10^7$ & $2.2 \times 10^5$ & $\sim 0.43$ & $\sim 26$\\
    FourierNetRGB$^2$ & $\bm{\uptheta}$ & $\sim1.6 \times 10^7$ & $1.1 \times 10^5$ & $\sim 0.47$ & $\sim 14$\\
    FourierUNetRGB$^1$ & $\bm{\uptheta}$ & $\sim7.1 \times 10^7$ & $2.5 \times 10^5$ & $\sim 3.3$ & $\sim 229$\\
    \midrule
    Le-ADMM-U$^2$ \cite{Monakhova_Yurtsever_Kuo_Antipa_Yanny_Waller_2019} & $\bm{\uptheta}$ & $\sim4.0 \times 10^7$ & - & - & -\\
    UNet$^2$ \cite{Monakhova_Yurtsever_Kuo_Antipa_Yanny_Waller_2019} & $\bm{\uptheta}$ & $\sim1.0 \times 10^8$ & - & - & -\\
    \bottomrule
\end{tabular}
\end{table}

\begin{table}[!h]
\centering
\caption{FourierNetRGB detailed architecture}
\label{tabfouriernetrgb}
\begin{tabular}{@{}lllll@{}}
    \toprule
    \textbf{Layer type} & \textbf{Kernel size} & \textbf{Stride} & \textbf{Notes} & \textbf{Shape (N, C, H, W)}\\
    \midrule
    FourierConv2D & (270, 480) & (2, 2) & - & (4, 3, 270, 480)\\
    LeakyReLU & - & - & slope: -0.01 & (4, 20, 270, 480)\\
    BatchNorm2D & - & - & - & (4, 20, 270, 480)\\
    Conv2D & (11, 11) & (1, 1) & - & (4, 64, 270, 480)\\
    BatchNorm2D & - & - & - & (4, 64, 270, 480)\\
    LeakyReLU & - & - & slope: -0.01 & (4, 64, 270, 480)\\
    Conv2D & (11, 11) & (1, 1) & - & (4, 64, 270, 480)\\
    BatchNorm2D & - & - & - & (4, 64, 270, 480)\\
    LeakyReLU & - & - & slope: -0.01 & (4, 64, 270, 480)\\
    Conv2D & (11, 11) & (1, 1) & - & (4, 3, 270, 480)\\
    ReLU & - & - & - & (4, 3, 270, 480)\\
    \bottomrule
\end{tabular}
\end{table}

\begin{table}[!h]
\centering
\caption{FourierUNetRGB detailed architecture}
\label{tabfourierunetrgb}
\begin{tabular}{@{}llp{25mm}lllp{28.5mm}@{}}
    \toprule
    \textbf{Scale} & \textbf{Repeat} & \textbf{Layer type} & \textbf{Kernel size} & \textbf{Stride} & \textbf{Notes} & \textbf{Shape (N, C, H, W)}\\
    \midrule
    1 & 1 & Multiscale\newline FourierConv2D\newline + ReLU\newline + BatchNorm2D & (270, 480) & (2, 2) & - & (4, 64, 270, 480)\\
    2 & & & (135, 240) & (2, 2) & & (4, 64, 135, 240)\\
    3 & & & (67, 120) & (2, 2) & & (4, 64, 67, 120)\\
    4 & & & (33, 60) & (2, 2) & & (4, 64, 33, 60)\\
    \midrule
    3 & 1 & Upsample2D & - & - & - & (4, 64, 67, 120)\\
    3 & 2 & Conv2D\newline + ReLU\newline + BatchNorm2D & (11, 11) & (1, 1) & - & (4, 64, 67, 120)\\
    2 & 1 & Upsample2D & - & - & - & (4, 64, 135, 240)\\
    2 & 2 & Conv2D\newline + ReLU\newline + BatchNorm2D & (11, 11) & (1, 1) & - & (4, 64, 135, 240)\\
    1 & 1 & Upsample2D & - & - & - & (4, 64, 270, 480)\\
    1 & 2 & Conv2D\newline + ReLU\newline + BatchNorm2D & (11, 11) & (1, 1) & - & (4, 64, 270, 480)\\
    1 & 1 & Conv2D\newline + ReLU & (1, 1) & (1, 1) & - & (4, 3, 270, 480)\\
    \bottomrule
\end{tabular}
\end{table}

\subsection{Details for FourierNets outperform UNets for engineering non-local optical encoders and 3D snapshot microscopy volume reconstruction}
\label{appendixnetworkcomparison}
\begin{figure}[h]
  \includegraphics[width=\linewidth]{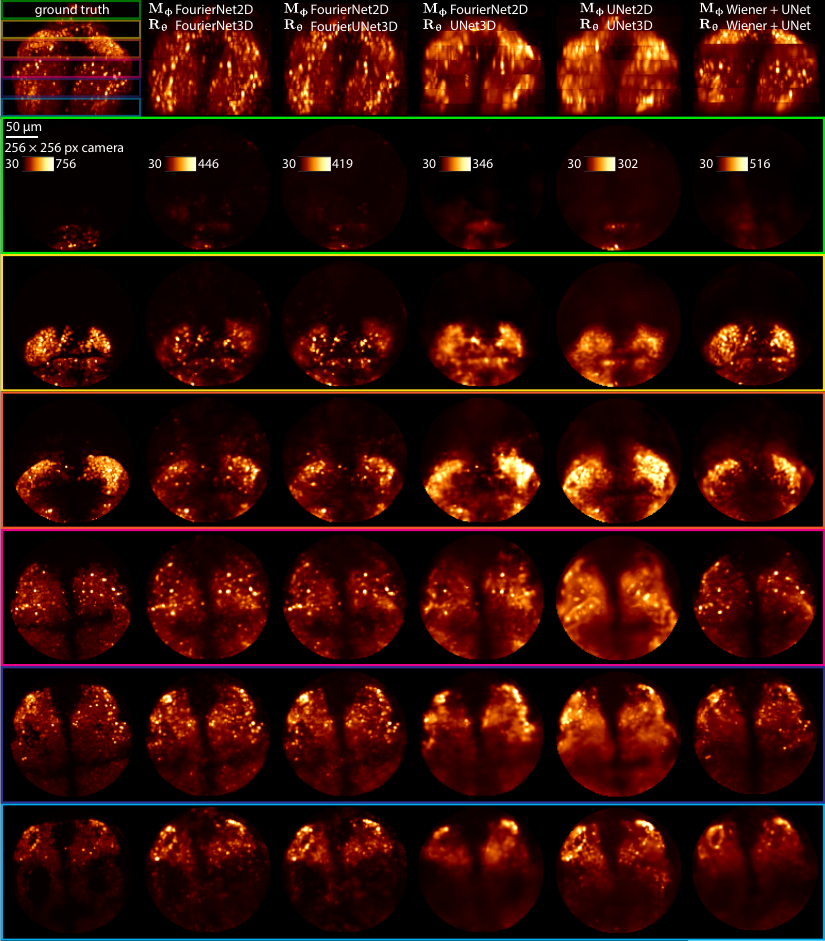}
  \caption{Slab views of a small example volume reconstruction, showing our methods (FourierNet/FourierUNet) do the best job of reconstructing throughout the volume. Note that the UNet reconstructions are blurry across all slabs, with few exceptions, while the Wiener + UNet reconstruction is not blurry in some slabs, but not consistent across all slabs. Colored boxes show which sample planes a particular slab comes from, corresponding to boxes in xz projection view at top. Annotation $\mathbf{M}_{\bm{\upphi}}$ shows which network architecture was used for phase mask optimization; annotation $\mathbf{R}_{\bm{\uptheta}}$ shows which architecture was used for reconstruction.}
  \label{figsupsmallslab}
\end{figure}

\begin{table}[!h]
\centering
\caption{FourierNet2D detailed architecture (1 per plane)}
\label{tabfouriernet2dsmall}
\begin{tabular}{@{}lllll@{}}
    \toprule
    \textbf{Layer type} & \textbf{Kernel size} & \textbf{Stride} & \textbf{Notes} & \textbf{Shape (C, D, H, W)}\\
    \midrule
    InputScaling & - & - & scale: 0.01 & (1, 1, 256, 256)\\
    FourierConv2D & (256, 256) & (2, 2) & - & (8, 1, 256, 256)\\
    LeakyReLU & - & - & slope: -0.01 & (8, 1, 256, 256)\\
    BatchNorm2D & - & - & - & (8, 1, 256, 256)\\
    Conv2D & (11, 11) & (1, 1) & - & (1, 1, 256, 256)\\
    ReLU & - & - & - & (1, 1, 256, 256)\\
    InputRescaling & - & - & scale: 0.01 & (1, 1, 256, 256)\\
    \bottomrule
\end{tabular}
\end{table}

\begin{table}[!h]
\centering
\caption{FourierNet3D detailed architecture (8 GPUs)}
\label{tabfouriernet3dsmall}
\begin{tabular}{@{}lllll@{}}
    \toprule
    \textbf{Layer type} & \textbf{Kernel size} & \textbf{Stride} & \textbf{Notes} & \textbf{Shape (C, D, H, W)}\\
    \midrule
    InputScaling & - & - & scale: 0.01 & (1, 1, 256, 256)\\
    FourierConv2D & (256, 256) & (2, 2) & - & (60, 1, 256, 256)\\
    LeakyReLU & - & - & slope: -0.01 & (60, 1, 256, 256)\\
    BatchNorm2D & - & - & - & (60, 1, 256, 256)\\
    Reshape2D3D & - & - & - & (5, 12, 256, 256)\\
    Conv3D & (11, 7, 7) & (1, 1, 1) & - & (5, 12, 256, 256)\\
    LeakyReLU & - & - & slope: -0.01 & (5, 12, 256, 256)\\
    BatchNorm3D & - & - & - & (5, 12, 256, 256)\\
    Conv3D & (11, 7, 7) & (1, 1, 1) & - & (1, 12, 256, 256)\\
    ReLU & - & - & - & (1, 12, 256, 256)\\
    InputRescaling & - & - & scale: 0.01 & (1, 12, 256, 256)\\
    \bottomrule
\end{tabular}
\end{table}

\begin{table}[!h]
\centering
\caption{FourierUNet3D detailed architecture (8 GPUs)}
\label{tabfourierunet3dsmall}
\begin{tabular}{@{}llp{25mm}lllp{28.5mm}@{}}
    \toprule
    \textbf{Scale} & \textbf{Repeat} & \textbf{Layer type} & \textbf{Kernel size} & \textbf{Stride} & \textbf{Notes} & \textbf{Shape (C, D, H, W)}\\
    \midrule
    1 & 1 & InputScaling & - & - & scale: 0.01 & (1, 1, 256, 256)\\
    1 & 1 & Multiscale\newline FourierConv2D\newline + ReLU\newline + BatchNorm2D & (256, 256) & (2, 2) & - & (60, 1, 256, 256)\\
    2 & & & (128, 128) & (2, 2) & & (60, 1, 128, 128)\\
    3 & & & (64, 64) & (2, 2) & & (60, 1, 64, 64)\\
    4 & & & (32, 32) & (2, 2) & & (60, 1, 32, 32)\\
    \midrule
    4 & 1 & Reshape2D3D & - & - & - & (5, 12, 32, 32)\\
    3 & 1 & Upsample2D & - & - & - & (5, 12, 64, 64)\\
    3 & 2 & Conv3D\newline + ReLU\newline + BatchNorm3D & (11, 7, 7) & (1, 1, 1) & - & (5, 12, 64, 64)\\
    2 & 1 & Upsample2D & - & - & - & (5, 12, 128, 128)\\
    2 & 2 & Conv3D\newline + ReLU\newline + BatchNorm3D & (11, 7, 7) & (1, 1, 1) & - & (5, 12, 128, 128)\\
    1 & 1 & Upsample2D & - & - & - & (5, 12, 256, 256)\\
    1 & 2 & Conv3D\newline + ReLU\newline + BatchNorm3D & (11, 7, 7) & (1, 1, 1) & - & (5, 12, 256, 256)\\
    1 & 1 & Conv3D\newline + ReLU & (1, 1, 1) & (1, 1, 1) & - & (1, 12, 256, 256)\\
    1 & 1 & InputRescaling & - & - & scale: 0.01 & (1, 12, 256, 256)\\
    \bottomrule
\end{tabular}
\end{table}

\begin{table}[!h]
\centering
\caption{UNet2D detailed architecture (1 per plane)}
\label{tabunet2dsmall}
\begin{tabular}{@{}llp{25mm}lllp{28.5mm}@{}}
    \toprule
    \textbf{Scale} & \textbf{Repeat} & \textbf{Layer type} & \textbf{Kernel size} & \textbf{Stride} & \textbf{Notes} & \textbf{Shape (C, D, H, W)}\\
    \midrule
    1 & 1 & InputScaling & - & - & scale: 0.01 & (1, 1, 256, 256)\\
    1 & 1 & Conv2D\newline + ReLU\newline + BatchNorm2D & (7, 7) & (1, 1) & - & (12, 1, 256, 256)\\
    1 & 1 & Conv2D\newline + ReLU\newline + BatchNorm2D & (7, 7) & (1, 1) & - & (24, 1, 256, 256)\\
    2 & 1 & MaxPool2D & (2, 2) & (2, 2) & - & (24, 1, 128, 128)\\
    2 & 2 & Conv2D\newline + ReLU\newline + BatchNorm2D & (7, 7) & (1, 1) & - & (24, 1, 128, 128)\\
    $n$ & 1 & MaxPool2D & (2, 2) & (2, 2) & - & (24, 1, $\frac{256}{2^{n - 1}}$, $\frac{256}{2^{n - 1}}$)\\
    $n$ & 2 & Conv2D\newline + ReLU\newline + BatchNorm2D & (7, 7) & (1, 1) & - & (24, 1, $\frac{256}{2^{n - 1}}$, $\frac{256}{2^{n - 1}}$)\\
    8 & 1 & MaxPool2D & (2, 2) & (2, 2) & - & (24, 1, 2, 2)\\
    8 & 2 & Conv2D\newline + ReLU\newline + BatchNorm2D & (7, 7) & (1, 1) & - & (24, 1, 2, 2)\\
    \midrule
    7 & 1 & Upsample2D & - & - & - & (24, 1, 4, 4)\\
    7 & 2 & Conv2D\newline + ReLU\newline + BatchNorm2D & (7, 7) & (1, 1) & - & (24, 1, 4, 4)\\
    $n$ & 1 & Upsample2D & - & - & - & (24, 1, $\frac{256}{2^{n - 1}}$, $\frac{256}{2^{n - 1}}$)\\
    $n$ & 2 & Conv2D\newline + ReLU\newline + BatchNorm2D & (7, 7) & (1, 1) & - & (24, 1, $\frac{256}{2^{n - 1}}$, $\frac{256}{2^{n - 1}}$)\\
    1 & 1 & Upsample2D & - & - & - & (24, 1, 256, 256)\\
    1 & 2 & Conv2D\newline + ReLU\newline + BatchNorm2D & (7, 7) & (1, 1) & - & (24, 1, 256, 256)\\
    1 & 1 & Conv2D\newline + ReLU & (1, 1) & (1, 1) & - & (1, 1, 256, 256)\\
    1 & 1 & InputRescaling & - & - & scale: 0.01 & (1, 1, 256, 256)\\
    \bottomrule
\end{tabular}
\end{table}

\begin{table}[!h]
\centering
\caption{UNet3D detailed architecture (8 GPUs)}
\label{tabunet3dsmall}
\begin{tabular}{@{}llp{25mm}lllp{28.5mm}@{}}
    \toprule
    \textbf{Scale} & \textbf{Repeat} & \textbf{Layer type} & \textbf{Kernel size} & \textbf{Stride} & \textbf{Notes} & \textbf{Shape (C, D, H, W)}\\
    \midrule
    1 & 1 & InputScaling & - & - & scale: 0.01 & (1, 1, 256, 256)\\
    1 & 1 & Conv2D\newline + ReLU\newline + BatchNorm2D & (7, 7) & (1, 1) & - & (30, 1, 256, 256)\\
    1 & 1 & Conv2D\newline + ReLU\newline + BatchNorm2D & (7, 7) & (1, 1) & - & (60, 1, 256, 256)\\
    2 & 1 & MaxPool2D & (2, 2) & (2, 2) & - & (60, 1, 128, 128)\\
    2 & 2 & Conv2D\newline + ReLU\newline + BatchNorm2D & (7, 7) & (1, 1) & - & (60, 1, 128, 128)\\
    3 & 1 & MaxPool2D & (2, 2) & (2, 2) & - & (60, 1, 64, 64)\\
    3 & 2 & Conv2D\newline + ReLU\newline + BatchNorm2D & (7, 7) & (1, 1) & - & (60, 1, 64, 64)\\
    4 & 1 & MaxPool2D & (2, 2) & (2, 2) & - & (60, 1, 32, 32)\\
    4 & 2 & Conv2D\newline + ReLU\newline + BatchNorm2D & (7, 7) & (1, 1) & - & (60, 1, 32, 32)\\
    \midrule
    4 & 1 & Reshape2D3D & - & - & - & (5, 12, 32, 32)\\
    3 & 1 & Upsample2D & - & - & - & (5, 12, 64, 64)\\
    3 & 2 & Conv3D\newline + ReLU\newline + BatchNorm3D & (11, 7, 7) & (1, 1, 1) & - & (5, 12, 64, 64)\\
    2 & 1 & Upsample2D & - & - & - & (5, 12, 128, 128)\\
    2 & 2 & Conv3D\newline + ReLU\newline + BatchNorm3D & (11, 7, 7) & (1, 1, 1) & - & (5, 12, 128, 128)\\
    1 & 1 & Upsample2D & - & - & - & (5, 12, 256, 256)\\
    1 & 2 & Conv3D\newline + ReLU\newline + BatchNorm3D & (11, 7, 7) & (1, 1, 1) & - & (5, 12, 256, 256)\\
    1 & 1 & Conv3D\newline + ReLU & (1, 1, 1) & (1, 1, 1) & - & (1, 12, 256, 256)\\
    1 & 1 & InputRescaling & - & - & scale: 0.01 & (1, 12, 256, 256)\\
    \bottomrule
\end{tabular}
\end{table}

\begin{table}[!h]
\centering
\caption{Wiener + UNet detailed architecture (8 GPUs)}
\label{tabwienersmall}
\begin{tabular}{@{}llp{25mm}llp{15mm}p{28mm}@{}}
    \toprule
    \textbf{Scale} & \textbf{Repeat} & \textbf{Layer type} & \textbf{Kernel size} & \textbf{Stride} & \textbf{Notes} & \textbf{Shape (C, D, H, W)}\\
    \midrule
    1 & 1 & InputScaling & - & - & scale: 0.01 & (1, 1, 256, 256)\\
    1 & 1 & WienerFilter & - & - & - & (12, 1, 256, 256)\\
    1 & 1 & Reshape2D3D & - & - & - & (1, 12, 256, 256)\\
    1 & 1 & Conv3D\newline + LeakyReLU\newline & (3, 3, 3) & (1, 1) & - & (32, 12, 256, 256)\\
    1 & 1 & Conv3D\newline + LeakyReLU\newline & (3, 3, 3) & (1, 1) & - & (32, 12, 256, 256)\\
    2 & 1 & Conv3D & (2, 2, 2) & (2, 2, 2) & - & (32, 6, 128, 128)\\
    2 & 2 & Conv3D\newline + LeakyReLU\newline & (3, 3, 3) & (1, 1) & - & (64, 6, 128, 128)\\
    3 & 1 & Conv3D & (2, 2, 2) & (2, 2, 2) & - & (64, 3, 64, 64)\\
    3 & 2 & Conv3D\newline + LeakyReLU\newline & (3, 3, 3) & (1, 1) & - & (128, 3, 64, 64)\\
    4 & 1 & Conv3D & (2, 2, 2) & (2, 2, 2) & - & (128, 1, 32, 32)\\
    4 & 2 & Conv3D\newline + LeakyReLU\newline & (3, 3, 3) & (1, 1) & - & (256, 1, 32, 32)\\
    5 & 1 & Conv3D & (2, 2, 2) & (2, 2, 2) & padding\newline(1, 0, 0) & (256, 1, 16, 16)\\
    5 & 2 & Conv3D\newline + LeakyReLU\newline & (3, 3, 3) & (1, 1) & - & (256, 1, 16, 16)\\
    \midrule
    4 & 1 & ConvTranspose3D & (1, 2, 2) & (2, 2, 2) & - & (256, 1, 32, 32)\\
    4 & 2 & Conv3D\newline + LeakyReLU\newline & (3, 3, 3) & (1, 1) & - & (128, 1, 32, 32)\\
    3 & 1 & ConvTranspose3D & (2, 2, 2) & (2, 2, 2) & output padding\newline(1, 0, 0) & (64, 3, 64, 64)\\
    3 & 2 & Conv3D\newline + LeakyReLU\newline & (3, 3, 3) & (1, 1) & - & (64, 3, 64, 64)\\
    2 & 1 & ConvTranspose3D & (2, 2, 2) & (2, 2, 2) & - & (64, 6, 128, 128)\\
    2 & 2 & Conv3D\newline + LeakyReLU\newline & (3, 3, 3) & (1, 1) & - & (32, 6, 128, 128)\\
    1 & 1 & ConvTranspose3D & (2, 2, 2) & (2, 2, 2) & - & (32, 12, 256, 256)\\
    1 & 2 & Conv3D\newline + LeakyReLU\newline & (3, 3, 3) & (1, 1) & - & (1, 12, 256, 256)\\
    1 & 1 & InputRescaling & - & - & scale: 0.01 & (1, 12, 256, 256)\\
    \bottomrule
\end{tabular}
\end{table}

For our experiments in Sections \ref{microscopeoptimizationcomparison} and \ref{networkcomparison}, we use 40 planes at 5$\mu$m resolution in z and therefore 40 reconstruction networks to train PSFs, except the Wiener + UNet model which is trained in a single stage. When training reconstruction networks only to produce the higher quality reconstructions, we use 96 planes at 1$\mu$m resolution in z (chosen so that the planes actually span 200 $\mu$m in z). Following \cite{Ikoma_Nguyen_Metzler_Peng_Wetzstein_2021}, the Wiener + UNet model is only trained in one stage (using knowledge of the current PSF, which the other methods do not receive), and is always trained on 96 planes. We train in both settings without any sparse planewise gradients, meaning we image and reconstruct all 40 or all 96 planes, respectively. We show details of all datasets used for training reconstructions in Table \ref{tabdatasets}.

We show the details of our FourierNet2D architecture for training PSFs in Table \ref{tabfouriernet2dsmall} and our FourierNet3D architecture for training reconstruction networks in Table \ref{tabfouriernet3dsmall}. We also show details for training times for both training PSFs and for training more powerful reconstruction networks in Table \ref{tabzdtrainingdetails}. We trained all networks for small $256 \times 256$ pixel experiments for the same number of iterations (more than necessary for PSFs to meaningfully converge)\footnote{Training times are approximate, and actual total time was longer due to checkpointing/snapshotting/validation of data and/or differences in load on the clusters being used.}.

The architecture of FourierUNet3D is 4 scales, with a cropping factor of 2 per scale in the encoding path and an upsampling factor of 2 in the decoding path. For each scale, we perform a Fourier convolution in the encoding path producing 480 feature maps, which are concatenated with the incoming feature maps of the decoding convolutions at the corresponding scale (just as in a normal UNet). In the decoding path, we use 3D convolutions with kernel size (3, 5, 5), producing 12 3D feature maps each. There are two such convolutions per scale. Note that this requires we reshape the 2D feature maps from the Fourier convolutions to 3D. This is followed by a 1x1 convolution producing the 3D reconstruction output. We show a diagram of this architecture in Figure \ref{figoverview}C, and details of this architecture in Table \ref{tabfourierunet3dsmall}.

For our UNet2D, each encoding convolution produced 24 feature maps (except the first scale, for which the first convolution produced 12 feature maps and the second convolution produced 24 feature maps). Each decoding convolution produced 24 feature maps, but took an input of 48 feature maps where 24 feature maps were concatenated from the corresponding encoding convolution at that scale. At the end of the UNet2D, a (1, 1) convolution reduced the 24 final feature maps to 1 feature map. This single feature map is interpreted as the final output of the network, i.e. the reconstructed plane. UNet2D requires many more feature maps per plane and more layers than FourierNet, because these are necessary in order for the network to be able to integrate information from a larger field of view. The effective field of view is $4,539 \times 4,539$ pixels. We show the details of our UNet2D architecture in Table \ref{tabunet2dsmall}.

The Wiener + UNet model first performs a Wiener deconvolution on the image using the PSF computed from the phase mask, then applies a UNet with a different 3D architecture, following that of \cite{Yanny_Yanny_Monakhova_Monakhova_Shuai_Waller_2022} which was designed to refine a Wiener filter. We show the details of this Wiener + UNet architecture in Table \ref{tabwienersmall}.

The architecture of the vanilla UNet3D is also 4 scales, with a max pooling factor of 2 per scale in the encoding path and an upsampling factor of 2 in the decoding path. Each scale of the encoding path produces 480 2D feature maps. These are concatenated to the incoming feature maps of the decoding convolutions at the corresponding scale, again with a reshape from 2D to 3D. Each scale of the decoding path produces 48 3D feature maps. Again, this is followed by a 1x1 convolution producing the 3D reconstruction output. All convolutions are in 3D with a kernel size of (5, 7, 7), with the $z$ dimension being ignored for the encoding path because the input is 2D. UNet3D has a global receptive field of $279 \times 279$ pixels. We show the details of our UNet3D architecture in Table \ref{tabunet3dsmall}.

\subsection{Details for engineered optical encoding depends on region of interest}
\label{appendixaperturecomparison}
\begin{figure}[!h]
  \includegraphics[width=\linewidth]{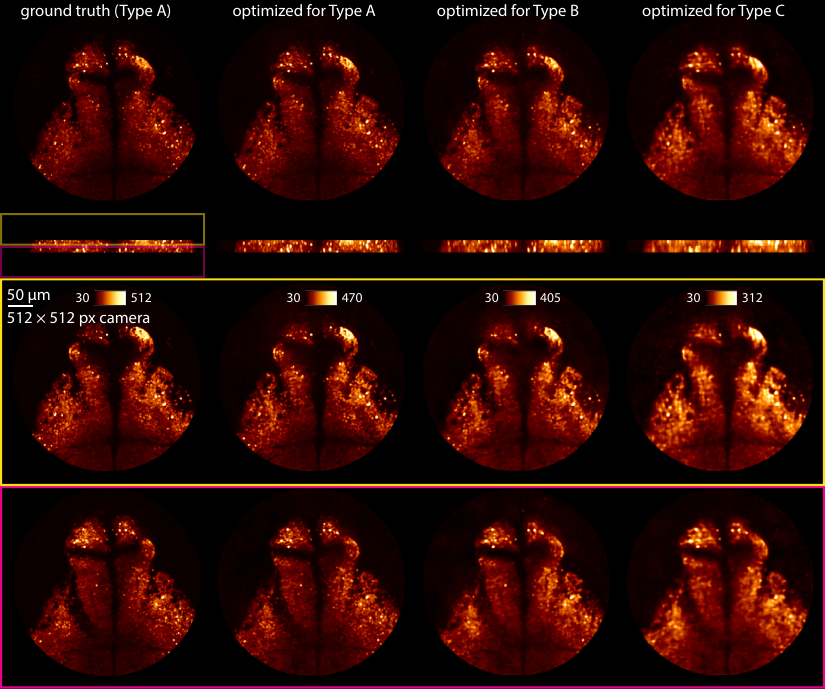}
  \caption{Slab views of an example Type A volume show that the phase mask optimized for Type A results in the best reconstructions. Note that the reconstruction with a phase mask optimized for Type A is almost identical to the ground truth, while the other phase masks create blurrier reconstructions. Slabs are xy max projections in thinner chunks as opposed to projecting through the entire volume. Colored boxes show which sample planes a particular slab comes from, corresponding to boxes in xz projection view at top.}
  \label{figsupa}
\end{figure}
\begin{figure}[!h]
  \includegraphics[width=\linewidth]{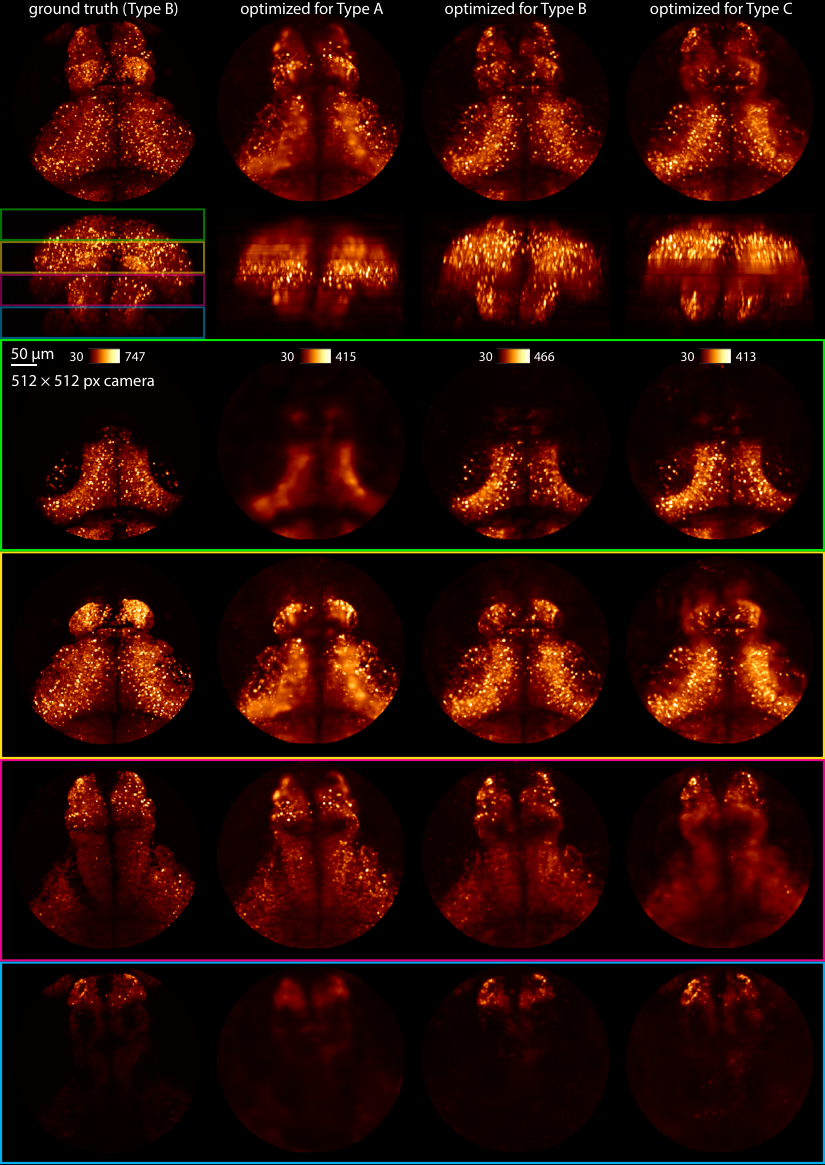}
  \caption{Slab views of an example Type B volume show that the phase mask optimized for Type B results in the best reconstructions; other phase masks result in blurrier reconstructions. Colored boxes show which sample planes a particular slab comes from, corresponding to boxes in xz projection view at top.}
  \label{figsupb}
\end{figure}
\begin{figure}[!h]
  \centering
  \includegraphics[width=0.975\linewidth]{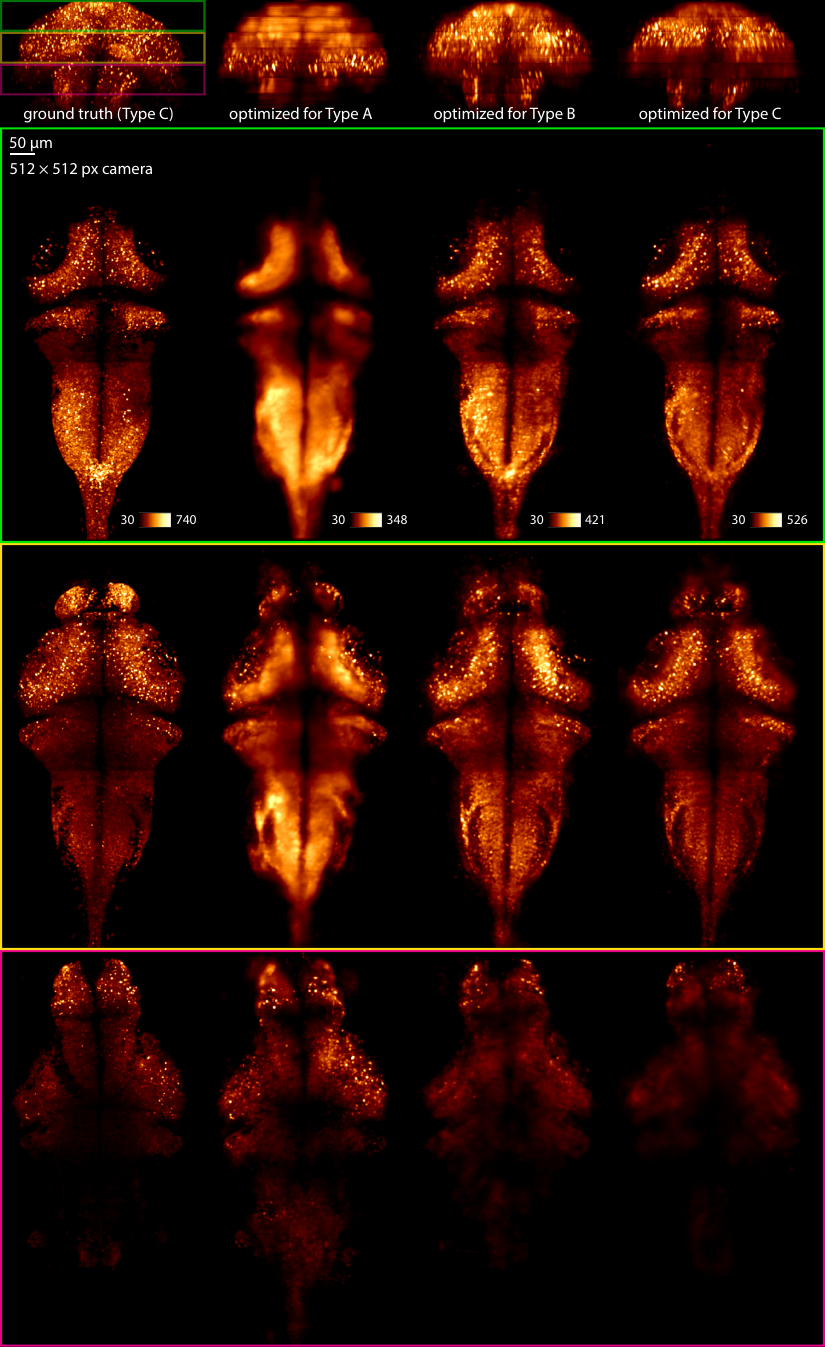}
  \caption{Slab views of an example Type C volume show that phase mask optimized for Type C provides most consistent reconstruction. Colored boxes have same meaning as Figures \ref{figsupa}, \ref{figsupb}.}
  \label{figsupc}
\end{figure}

\begin{table}[!h]
\centering
\caption{FourierNet2D detailed architecture (1 per plane)}
\label{tabfouriernet2dlarge}
\begin{tabular}{@{}lllll@{}}
    \toprule
    \textbf{Layer type} & \textbf{Kernel size} & \textbf{Stride} & \textbf{Notes} & \textbf{Shape (C, D, H, W)}\\
    \midrule
    InputScaling & - & - & scale: 0.01 & (1, 1, 512, 512)\\
    FourierConv2D & (512, 512) & (2, 2) & - & (5, 1, 512, 512)\\
    LeakyReLU & - & - & slope: -0.01 & (5, 1, 512, 512)\\
    BatchNorm2D & - & - & - & (5, 1, 512, 512)\\
    Conv2D & (11, 11) & (1, 1) & - & (1, 1, 512, 512)\\
    ReLU & - & - & - & (1, 1, 512, 512)\\
    InputRescaling & - & - & scale: 0.01 & (1, 1, 512, 512)\\
    \bottomrule
\end{tabular}
\end{table}

\begin{table}[!h]
\centering
\caption{FourierNet3D detailed architecture (8 GPUs)}
\label{tabfouriernet3dlarge}
\begin{tabular}{@{}lllll@{}}
    \toprule
    \textbf{Layer type} & \textbf{Kernel size} & \textbf{Stride} & \textbf{Notes} & \textbf{Shape (C, D, H, W)}\\
    \midrule
    InputScaling & - & - & scale: 0.01 & (1, 1, 512, 512)\\
    FourierConv2D & (512, 512) & (2, 2) & - & (80, 1, 512, 512)\\
    LeakyReLU & - & - & slope: -0.01 & (80, 1, 512, 512)\\
    BatchNorm2D & - & - & - & (80, 1, 512, 512)\\
    Reshape2D3D & - & - & - & (5, 16, 512, 512)\\
    Conv3D & (11, 7, 7) & (1, 1, 1) & - & (5, 16, 512, 512)\\
    LeakyReLU & - & - & slope: -0.01 & (5, 16, 512, 512)\\
    BatchNorm3D & - & - & - & (5, 16, 512, 512)\\
    Conv3D & (11, 7, 7) & (1, 1, 1) & - & (1, 16, 512, 512)\\
    ReLU & - & - & - & (1, 16, 512, 512)\\
    InputRescaling & - & - & scale: 0.01 & (1, 16, 512, 512)\\
    \bottomrule
\end{tabular}
\end{table}

For our experiments in Section \ref{aperturecomparison}, we use 64 planes at 1$\mu$m resolution in z and therefore 64 reconstruction networks to train PSFs. When training reconstruction networks only to produce the higher quality reconstructions, we use 128 planes at 1$\mu$m resolution in z (chosen so that the planes actually span 250 $\mu$m in z). We train in the reconstruction only setting without any sparse planewise gradients, meaning we image and reconstruct all 128 planes. However, when training a PSF we image and reconstruct 40 planes at a time with gradient per iteration (spread across 8 GPUs). These 40 planes are chosen randomly at every iteration from the 64 total possible planes, making potentially separate draws of planes for imaging and reconstruction. We show details of all datasets used for training reconstructions in Table \ref{tabdatasets}.

We show the details of our FourierNet2D architecture for training PSFs at the larger field of view in Type A, B, C in Table \ref{tabfouriernet2dlarge} and our FourierNet3D architecture for training reconstruction networks at the larger field of view in Type A, B, C in Table \ref{tabfouriernet3dlarge}. There are no other networks used for these larger field of view experiments. We also show details for training times for both training PSFs and for training more powerful reconstruction networks in Table \ref{tabzabctrainingdetails}. All PSFs in these networks were trained for the same number of iterations. However, reconstruction networks for some of these experiments were only trained for as long as necessary to converge (with some exceptions where we attempted longer training to check for performance gains with long training periods). Generally, we observed that performance for such reconstruction networks does not meaningfully change with many more iterations of training\footnote{Training times are approximate, and actual total time was longer due to checkpointing/snapshotting/validation of data and/or differences in load on the clusters being used.}.

\subsection{Details for engineered optical encoders implemented on a programmable microscope}
\label{appendixmeasurement}
The experimental data presented in this manuscript was acquired with a prototype programmable microscope. Light was collected by a 16X, 0.8 NA microscope objective (N16XLWD-PF, Nikon) and relayed onto a conjugate image plane by a 200 mm tube lens (TL200CLS2, Thorlabs). A polarizing beam splitter (PBS251, Thorlabs) transmitted horizontally polarised light which was relayed onto a spatial light modulator (P1920-532, Meadowlark) positioned in a conjugate pupil plane and used to modulate the phase of the beam with the optimized phase mask. The modulated light was imaged onto an sCMOS camera (Orca-Flash4.0 C11440, Hamamatsu). The total magnification between the focal plane of the objective and the sensor plane was 18.35X, resulting in an object space pixel size of 0.354 µm. To experimentally characterise the optical encoder (point spread function), an artificial point source was generated by focusing a collimated laser diode (532 nm, CPS532, Thorlabs) to a diffraction limited spot using a second, higher NA, microscope objective (60X 0.9 NA LUMPLFLN60XW, Olympus). The artificial point source was displaced about the focal plane of the primary microscope objective in 5 µm steps over a total range of 250 $\mu$m. 100 images were acquired at each plane, averaged, and dark frame subtracted.

The phase mask (microscope parameters $\bm{\upphi}$) was optimized for this spatial light modulator (SLM) by choosing a number of pixels for the phase mask such that the desired pupil size would fit on the physical pixels of the SLM. In order to simulate high frequencies accurately, we upsample this phase mask to the number of pixels used for all our other simulations. This phase mask was optimized for Type B samples.

For visualization in Figure \ref{figmeasuredpsf}, we clipped any values of the measured optical encoder that were below 0 after dark frame subtraction to 0, then simulated imaging. We simulated the optical encoder using the same wavelength as the laser point source (532 nm) for the measured optical encoder. For both the simulated and measured optical encoders, we scale the values to range from 0 to 1. We then simulated imaging using the same sample from our Type B dataset. The simulation for the measured optical encoder used a Type B sample interpolated to a resolution that matched the Orca-Flash camera.

\end{document}